\documentclass[aps, prl, superscriptaddress, groupedaddress, showkeys, showpacs, nofootinbib]{revtex4-2}
\usepackage{epsfig}
\usepackage{amssymb}
\usepackage{amsmath}
\usepackage{amsfonts}
\usepackage{graphicx}
\usepackage{mathrsfs}
\usepackage[dvipsnames]{xcolor}
\usepackage{multirow}
\usepackage{lipsum,appendix}
\usepackage{booktabs}
\usepackage{braket}
\usepackage{hyperref}
\usepackage[table]{xcolor}
\usepackage{longtable}
\usepackage{colortbl}
\usepackage{tikz}
\usepackage[utf8x]{inputenc}

\definecolor{type1}{RGB}{255, 49, 49}
\definecolor{type2}{RGB}{0, 74, 173}
\definecolor{type3}{RGB}{193, 255, 114}
\definecolor{type4}{RGB}{255, 222, 89}
\definecolor{type5}{RGB}{255, 145, 77}
\definecolor{type6}{RGB}{0, 191, 99}
\definecolor{type7}{RGB}{255, 102, 196}
\definecolor{type8}{RGB}{203, 108, 230}
\definecolor{type9}{RGB}{12, 192, 223}
\definecolor{type10}{RGB}{166, 166, 166}
\definecolor{type11}{RGB}{140, 82, 255}
\definecolor{type12}{RGB}{82, 113, 255}

\newcommand{\fn}{{\mathfrak{n}}}

\newcommand{\fz}{\mathfrak{z}}

\newcommand{\fK}{\mathfrak{K}}

\newcommand{\bA}{\mathbf{A}}

\newcommand{\bC}{\mathbf{C}}

\newcommand{\bR}{\mathbf{R}}

\newcommand{\bT}{\mathbf{T}}

\newcommand{\cB}{\mathcal{B}}

\newcommand{\cD}{\mathcal{D}}
\newcommand{\cH}{\mathcal{H}}
\newcommand{\cE}{\mathcal{E}}

\newcommand{\cJ}{\mathcal{J}}

\newcommand{\be}{\begin{equation}}
\newcommand{\ee}{\end{equation}}
\newcommand{\bea}{\begin{eqnarray}}
\newcommand{\eea}{\end{eqnarray}}

\newcommand{\ed}{\end{document}}

\newcommand{\bi}{\begin{itemize}}
\newcommand{\ei}{\end{itemize}}

\newcommand{\bce}{\begin{center}}
\newcommand{\ece}{\end{center}}

\newcommand{\sE}{\mathscr{E}}

\begin{document}

\title{Exploring Spectral Singularities and Topological Lasers in $\mathcal{PT}$-Symmetric Weyl Semimetals}

\author{Arda Sevin\c{c}}\email{arda.sevinc@ogr.iu.edu.tr}
\affiliation{Department of Physics, Istanbul University, 34134, Vezneciler, Istanbul, Türkiye}
\author{Rama Alassadi}\email{rama5552010@hotmail.com}
\affiliation{Institute of Graduate Studies in Science, Istanbul University, Istanbul 34134, Türkiye}
\author{Mustafa Sarısaman}\email{m.sarisaman@mia.edu.tr}\affiliation{Department of Physics, Istanbul University, 34134, Vezneciler,
Istanbul, Türkiye}
\affiliation{National Intelligence Academy, Institute of Engineering and Science, Ankara, Türkiye}

\begin{abstract}


This paper investigates the unique properties of $\mathcal{PT}$-symmetric Topological Weyl Semimetals (TWS) within the framework of non-Hermitian physics, focusing on their potential for generating topological lasers. By exploring the role of spectral singularities and their relationship to exceptional points, we examine how these materials, characterized by Weyl nodes and topologically protected surface states, can support novel optical phenomena such as unidirectional propagation and enhanced lasing. Through a theoretical model based on the transfer matrix approach, we reveal how the interplay between the $\mathcal{PT}$ symmetry and the axion term introduces new dynamics, leading to 12 distinct topological laser configurations. The study also investigates the impact of the $\theta$-term on spectral singularities, showing how it quantizes the system's gain values and influences the topological properties of the lasers. By applying our model to the TaAs material, a known Weyl semimetal, we uncover previously unreported effects, demonstrating the potential of $\mathcal{PT}$-symmetric TWS materials for advanced optoelectronic applications. We show that the axion-induced cyclotron-like Hall current in a $\mathcal{PT}$-symmetric TWS medium, revealing its topological characteristics and distinct flow patterns in the gain and loss regions, which serve as indicators of the system's topological symmetry. Our findings open new avenues for the development of robust, tunable, and efficient topological lasers with applications in quantum information processing and beyond.

\end{abstract}

\vspace{2mm}  
\vspace{2mm}
\maketitle

\section{Introduction}

Topology is being evaluated as a distinct domain within the discipline of mathematics that deals with geometric objects preserving its properties under continuous deformation – twisting, stretching without breaking or irreversibly changing its geometrical features. 
It's fascinating to discover that abstract mathematical idea of topology has real-world physical counterparts and possible practical applications \cite{topphys1, topphys2}. Recently, topological materials have been considered novel and promising in areas such as quantum matter, photonics, and electronics \cite{topphase1, topphase2, topphase3}. This class of materials is distinguished by its unique electronic and mechanical properties, which arise from their non-trivial topological features. As a result, considerable experimental and theoretical work is being carried out to broaden and enhance their practical applications \cite{topmats1, topmats2, topmats3, topmats4, topmats5, topmats6, topmats7, topmats8, topmats9, topmats10, topmats11, topmats12, topmats13}. Semimetals, a subgroup within this class, possess an electronic band structure that lies between those of metals and non-metals. Unlike metals and semiconductors, semimetals are characterized by the valence and conduction bands slightly intersecting at the Fermi energy level \cite{topmats3, topweyl1, topweyl2, topweyl3, topweyl4, topweyl5, topweyl6, topweyl7, topweyl8, topweyl9, topweyl10, topweyl11, topweyl12, topweyl13, topweyl14}. 

There are two opinions to explain the definition of band inversion in literature \cite{Narang, Takane}. Band inversion stems from element’s energy and orbital characteristics, spin-orbit interaction (SOC), or both simultaneously. As mentioned above, semimetals have a slightly intersecting band. However, in the case of heavy elements, the s and p or d orbitals interchange their energies; resulting in overlapping – “inverted”- bands. The other opinion suggests that this inversion happens due to the SOC (spin-orbit coupling). The points of intersection between s and p or d orbitals result in a line denoted as the nodal line. The topological characteristic emerges as a result of phase transition due to the spin and orbit coupling around the nodal line and results in the complete separation such as in Topological insulators, or separation around nodal points (nodes) such as Weyl and Dirac semimetals \cite{Singh2023, topmats3}.

In the burgeoning field of topological matter, Weyl semimetals stand out for their extraordinary electronic properties and rich topological features \cite{topweyl1, topweyl2, topweyl3, topweyl4, topweyl5, topweyl6, topweyl7, topweyl8, topweyl9, topweyl10, topweyl11, topweyl12, topweyl13, topweyl14, weylopt1, weylopt2, weylopt3, weylopt4, weylopt5, weylopt6, weylopt7, weylopt8, weylopt9, weylopt10, weylopt11, weylopt12, weylopt13, weylopt14, weylopt16, topweyl11}. These materials are distinguished by the presence of Weyl nodes—points in momentum space where conduction and valence bands touch, leading to intriguing phenomena such as Fermi arcs and chiral anomalies \cite{topweyl14, wfc2, wfc3, wfc4, wfc5, wfc6, wfc7, wfc8}. However, the exploration of Weyl semimetals has recently expanded beyond conventional boundaries to include systems with  $\mathcal{PT}$ symmetry (parity-time symmetry), which introduces a new layer of complexity and potential. These materials possess a variety of distinctive properties that make them highly suitable for advanced applications. Their potential spans numerous technologies, including quantum computing, electronics, spintronics, thermoelectrics, and photonics, among others \cite{uws1, uws2, weylopt8, uws4, uws5, uws6, weylopt4, uws8, uws9}. As our understanding of these materials expands and fabrication methods advance, the range of possible applications is expected to increase, paving the way for new technological innovations.

\begin{figure}
    \begin{center}
    \includegraphics[scale=.050]{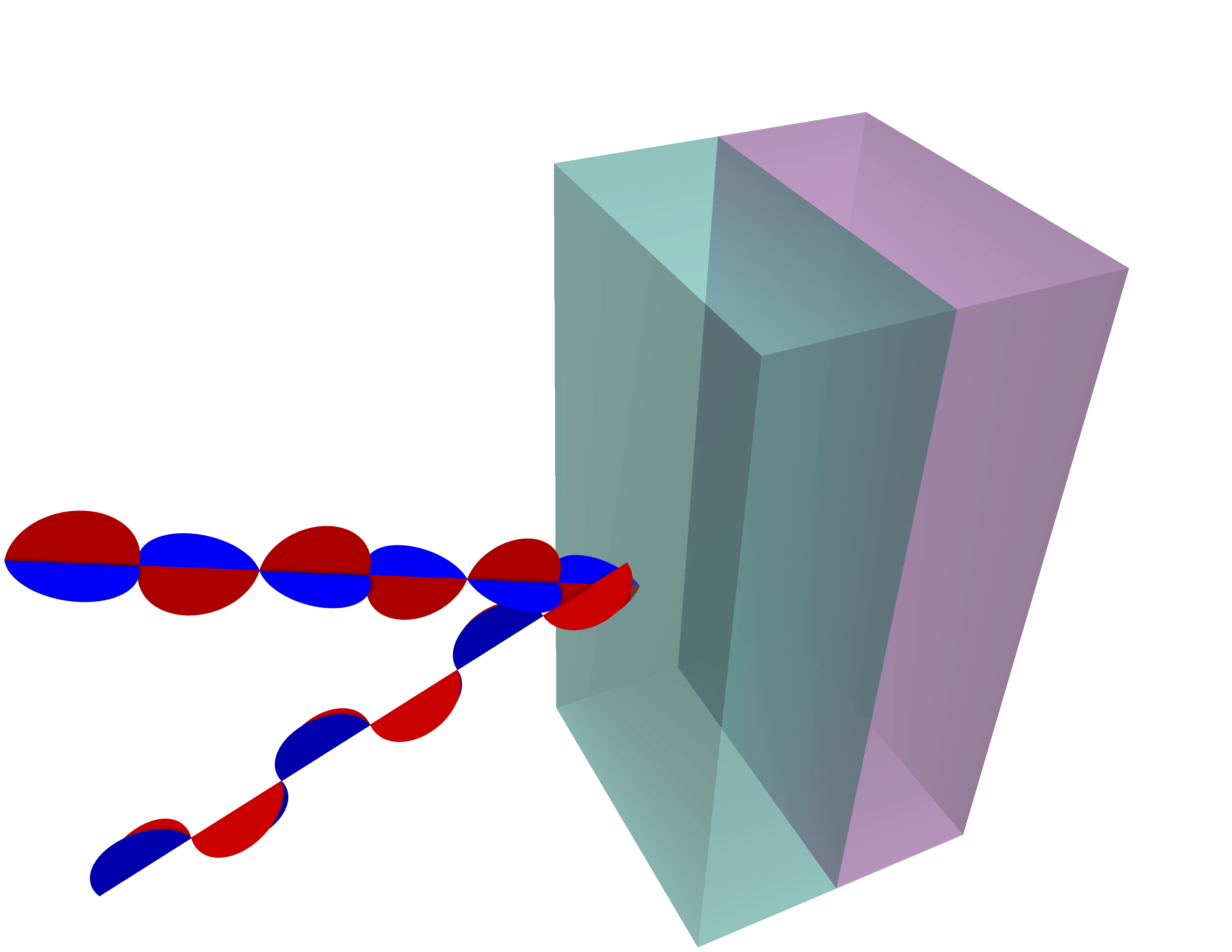}
    \includegraphics[scale=.050]{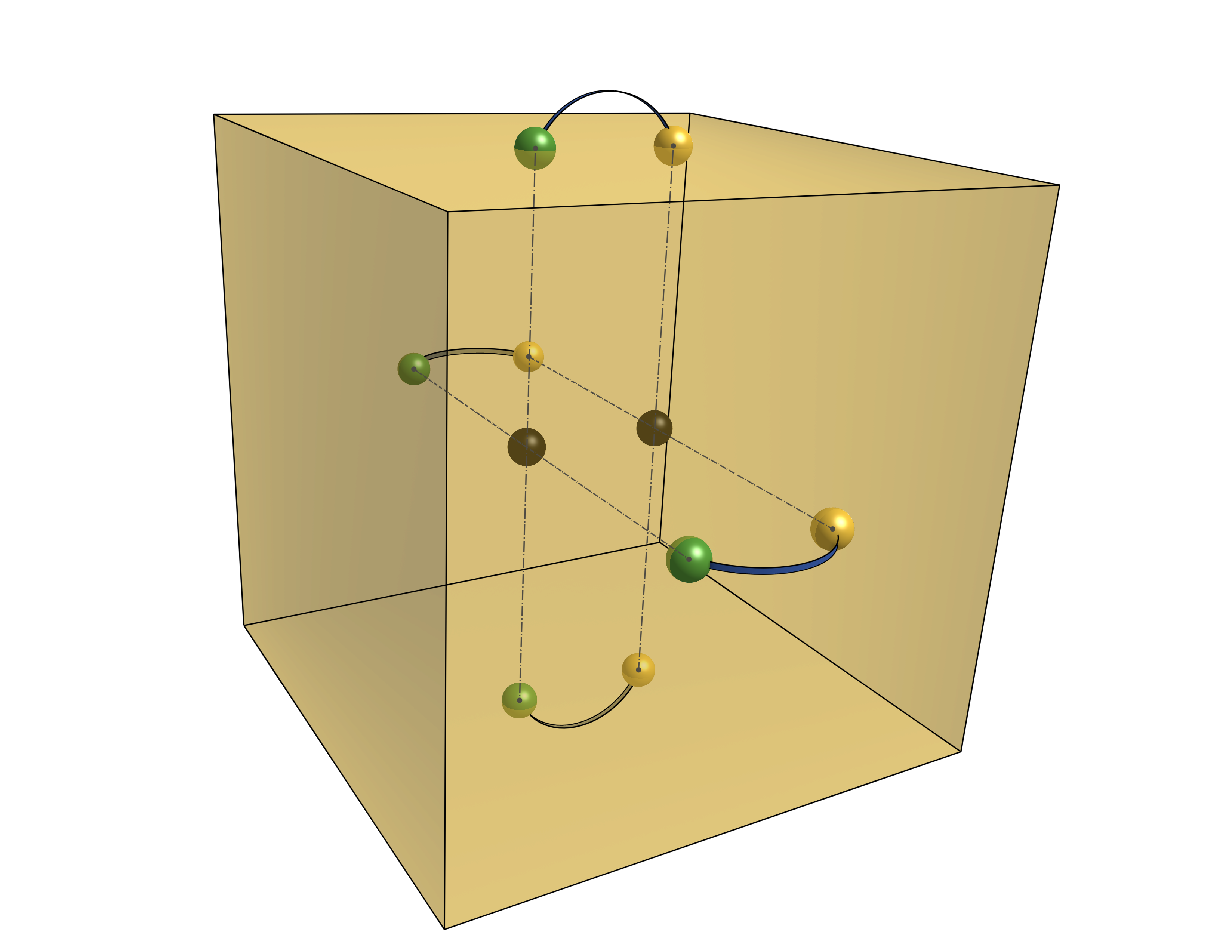}
    \caption{(Color online) The TE mode configuration for the interaction of an electromagnetic wave with the Weyl semimetal slab (Left panel). The wave is emitted on the slab by an angle $\phi$ which is measured from the normal to the surface, and direction of the polarization is rotated by an angle of $\phi_F$ and $\phi_K$ inside and outside of the slab respectively. On the right panel, Fermi arcs due to Weyl nodes for an individual slab are presented.}
    \label{fig1}
    \end{center}
    \end{figure}

In this study, we investigate a $\mathcal{PT}$-symmetric topological Weyl semimetal (TWS) and explore its non-Hermitian effects, particularly in the context of electromagnetic wave scattering, to propose new classes of topological lasers not yet reported in the literature. The topological nature of the material arises from a phase transition that introduces an axion term, endowing the system with its nontrivial topological properties \cite{axion1, axion2, axion3}.  Although significant progress has been made in understanding the behavior of TWSs, their optical interactions and full topological implications remain insufficiently explored \cite{weylopt1, weylopt2, weylopt3, weylopt4, weylopt5, weylopt6, weylopt7, weylopt8, weylopt9, weylopt10, weylopt11, weylopt12, weylopt13, weylopt14, weylopt16, topweyl11}. Our aim is to address this gap by studying $\mathcal{PT}$-symmetric TWS materials. We develop a theoretical model for a $\mathcal{PT}$-symmetric TWS laser using the transfer matrix approach. We demonstrate that the topological protection of surface states leads to 12 distinct topological laser types that respect $\mathcal{PT}$ symmetry, with quantized gain values required to meet the laser threshold condition\footnote{The term laser generally refers to the phenomenon in which a wave entering a gain medium increases in amplitude as it circulates within the medium, eventually reaching a certain threshold and producing only outgoing waves. For lasing to occur, two key conditions must be met: the presence of a gain medium that amplifies the wave amplitude, and the establishment of appropriate conditions that allow the wave to remain within this gain medium, thereby enabling the laser threshold condition to be reached. The analyses presented in this study are built upon these fundamental principles of the laser concept.}. The configuration, shown in Fig.~\ref{fig1}, consists of a TWS material with Weyl nodes aligned along the $z$-axis.

$\mathcal{PT}$ symmetry, which combines parity ($\mathcal{P}$) and time-reversal ($\mathcal{T}$) symmetry, plays a crucial role in the study of non-Hermitian systems, particularly in topological materials like $\mathcal{PT}$-symmetric topological Weyl semimetals \cite{PTS, PTS1, PTS2, PTS3, PTS4, PTS5, PTS6, PTS7, PTS8, PTS9, PTS10, PTS11, PTS12, PTS13, PTS14, PTS15}.  Unlike conventional Hermitian systems, where symmetries ensure the conservation of probability and lead to real eigenvalues, non-Hermitian systems with $\mathcal{PT}$ symmetry can exhibit unique phenomena such as the coalescence of eigenstates at exceptional points and the formation of robust edge states immune to perturbations \cite{PTS, bender, ijgmmp-2010, longhi4, longhi3, nonhermit1, nonhermit2, nonhermit3, nonhermit4, nonhermit5, nonhermit6, nonhermit7, nonhermit8, nonhermit9, nonhermit10, nonhermit11, nonhermit12, nonhermit13}. In these systems, the standard principles of quantum mechanics are altered, giving rise to new effects such as exceptional points, unidirectional light propagation, and enhanced laser performance \cite{pra-2012a, Oktay2020, CPA, lastpaper, CPA-8, CPA-9, pra-2011a, antilaser1, antilaser2, antilaser2-1, antilaser2-2, antilaser2-3, antilaser3, antilaser4, antilaser5, prl-2009, pra-2017a, cpa3, sevval, hamed2020, sarisaman2019, grapheneenergy, sar1, sarisaman2019}. This symmetry allows for the design of materials that are not only topologically protected but also capable of supporting unusual physical effects like unidirectional propagation or topologically protected lasing. In the context of TWS materials, $\mathcal{PT}$ symmetry enables the stabilization of surface states that would otherwise be unstable in non-Hermitian systems, offering new opportunities for controlling light-matter interactions, developing novel optoelectronic devices, and realizing advanced topological phases with potential applications in quantum information processing, sensing and lasing.   

Spectral singularities—a striking phenomenon where the density of states exhibits a peak of infinite height—have emerged as a critical topic in the study of non-Hermitian systems. These singularities are associated with exceptional points, where eigenvalues and eigenvectors coalesce, and they manifest in various physical contexts, from optics to quantum mechanics. In this context, the spectral singularities of an optical system, in case of scattering, are associated with the emergence of states where the reflection and transmission amplitudes diverge for the real values of $k$ in the physical system \cite{prl-2009, CPA, lastpaper, pra-2011a, naimark, naimark-1, p123, pra-2012a}. This leads to the occurrence of zero-width resonances and the laser threshold state, as it typically results in purely outgoing waves \cite{prl-2009}. Such behavior is a natural outcome of non-Hermitian physics, in contrast to conventional lasers. In recent years, significant progress has been made in exploring various unknown aspects of new phenomena within non-Hermitian physics, and a wealth of fascinating studies are currently underway \cite{nonhermit1, nonhermit2, nonhermit3, nonhermit4, nonhermit5, nonhermit6, nonhermit7, nonhermit8, nonhermit9, nonhermit10, nonhermit11, nonhermit12, nonhermit13}. Thus, non-Hermitian physics is essential for understanding the unique properties of topological materials \cite{Oktay2020, hamed2020, sarisaman2019, CPA-9}, and it serves as the primary motivation for our work. Studying topological systems within the context of non-Hermitian physics presents an exciting and innovative approach. By incorporating Weyl semimetals into these framework, researchers are discovering new ways to design lasers with exceptional efficiency, tunability, and robustness. In the realm of $\mathcal{PT}$-symmetric systems, these singularities gain additional significance, offering insights into the interplay between symmetry, topology, and spectral properties.

This paper explores the intriguing intersection of $\mathcal{PT}$ symmetry and Weyl semimetals through the lens of spectral singularities. We will begin by introducing the foundational concepts of $\mathcal{PT}$ symmetry and Weyl semimetals, detailing how these materials exhibit unique topological characteristics and how $\mathcal{PT}$ symmetry influences their electronic behavior. Next, we will delve into the concept of spectral singularities, elucidating their origins, implications, and how they manifest in the context of $\mathcal{PT}$-symmetric Weyl semimetals. By focusing on spectral singularities, we aim to illuminate how these phenomena impact the electronic structure and physical properties of $\mathcal{PT}$-symmetric topological Weyl semimetals. Our discussion will reveal the potential of these materials to exhibit novel spectral features and their implications for future research and technological applications in topological and non-Hermitian systems.

Given the growing interest in this field and the fact that our topological material of focus has an optically active structure, we will examine how it interacts with electromagnetic waves. Recently, the discovery of Kerr/Faraday effects in Weyl semimetals has shown how important it is to study these interactions.  We know that Kerr and Faraday rotations in the Weyl semimetals lead to an increase in the size of the system, such that one encounters an additional computational challenge in the structure\cite{Oktay2020}. However, this challenge also reveals deeper insights into the system. Through our study, we uncovered new and previously unknown aspects of TWS by exploring these complexities. To address this, we designed our system so that the Kerr/Faraday effect results in a 4x4 transfer matrix, leading to 12 distinct lasing configurations, some of which have topologically robust features. We aim to explore the topological effects in our system by generating waves in the TE mode, which will allow us to study the system’s topological characteristics. It is well-established that the topological properties of such systems are governed by the $\theta$-term \cite{axion2, theta1}, where in our case, the term is simply $\mathtt{b}$, distance between Weyl nodes in the bulk structure. In axion electrodynamics, the $\theta$-term introduces a topological coupling between electric and magnetic fields, leading to novel magnetoelectric effects, see Appendix B how this coupling occurs in our case. This term plays a crucial role in systems such as topological insulators and Weyl semimetals, where it gives rise to observable phenomena like the anomalous Hall effect and optical activity.

To understand how the $\theta$-term affects the system's topological properties, we will analyze the scattering behavior of $\mathcal{PT}$-symmetric TWS, identify spectral singularities, and investigate how the $\theta$-term influences these singularities. Spectral singularities are points where the system’s continuous spectrum has exceptional characteristics\cite{naimark, naimark-1, p123}. Thus, the interaction of TWS with electromagnetic waves can be viewed as a non-Hermitian scattering problem in electromagnetic theory\cite{Oktay2020}.

Our work proceeds as follows: First, we calculated the transfer matrix through boundary conditions by solving Maxwell's equations with an axion term, which are specific to Weyl semimetal, for the TE mode configuration. The transfer matrix allows us to calculate spectral singularities. By calculating the spectral singularities in this way, in the last part, we determined the effect of the $\theta$-term on the spectral singularities by using the TaAs material, which has been experimentally proven to be suitable for the Weyl semimetal phase \cite{taas1, topweyl12, taas3, taas4, taas5, taas6, taasp1, taasp2, taasp3}. We obtained novel results in our analysis. Accordingly, we show that the presence of the $\theta$-term and $\mathcal{PT}$-symmetry  significantly reduce the gain value in the system. Again, it has been manifestly shown that it is topologically quantized by degenerating the spectral singularity points in the system. This result is very important and has been noticed for the first time in the literature. We finally find out the axion induced current present inside the TWS medium. These currents, arising from the axion term $\theta$, exhibit cyclotron-like Hall current patterns confined to the $xy$-plane, with distinct behaviors in the gain and loss regions.  The results of this study shows that 12 different topological laser types can be created due to the Kerr/Faraday effect in the TWS material and under what conditions these lasers can exist.

\section{Interaction of Electromagnetic Waves with a Single TWS Slab}
\label{S2}

Since the transfer matrix approach will be adopted in our study, we will use the associative property, which is the practical and best-handy property of the transfer matrix. Using this feature, the transfer matrix of a multi-layered holistic system placed side by side can be obtained by multiplying the individual transfer matrices. Thus, the total transfer matrix of a $\mathcal{PT}$-symmetric system is found by multiplying the individual transfer matrices of the components that make up the system. An important advantage of $\mathcal{PT}$ symmetry is that the components of the system can be given in terms of each other. Accordingly, if the potential expression of one component is $V(\mathbf{z})$, the other will be $V^{*}(-\mathbf{z})$. This necessitates that if one component of the system is gain, the other must be loss with the same amount. Therefore, if the transfer matrix of one component can be found saliently, it will be sufficient to simply substitute loss for gain in the other.

\begin{figure}
    \begin{center}
    \includegraphics[scale=0.60]{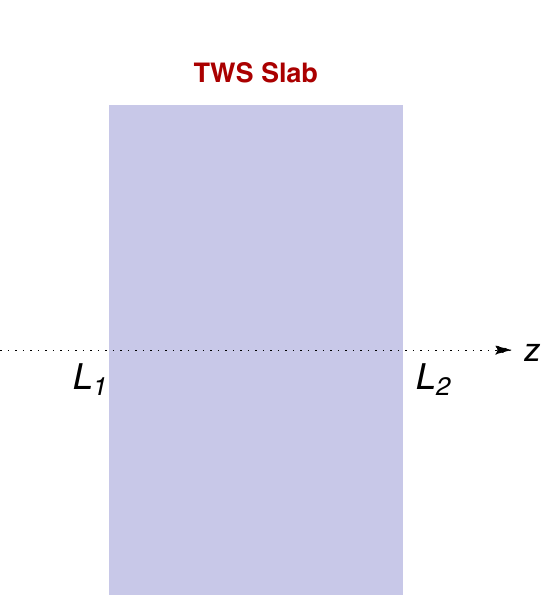}
    \caption{(Color online) Configurations of a single layer optically active TWS slab (side view).}
    \label{fig1tws}
    \end{center}
    \end{figure}

Consider a linear, homogeneous and planar TWS slab aligned in $z$-axis such that it consists of optically active gain/loss component. In the $x-$ and $y-$ directions, the system maintains the same homogeneous and isotropic content and is in the desired length. Therefore, although our system appears to be 1-dimensional, it is actually a 3-dimensional TWS system. Another issue that we should mention is that although the system may exhibit behavioral anomalies due to uncontrolled temperature, disorder or impurity differences, we will assume that such changes do not occur in our study. The importance of these effects is the subject of another study. Due to our geometric and analytical concerns, let the TWS slab be confined between the boundaries $L_1$ and $L_2$ along the $z$-axis as shown in the Fig.~(\ref{fig1tws}). The reason for positioning the material content along the $z$-axis is that the axion term (i.e. $\mathtt{b}$-term) of TWS is present exclusively along this axis.

To understand wave propagation in a TWS environment, it is essential to consider the role of the $\theta$-term derived from the material properties. This $\theta$-term, known as magneto-electric polarizability, is generally associated with the Berry phase and Chern number. In the context of electrodynamic interactions, it is addressed by axion electrodynamics. Typically, ordinary insulators have $\theta = 0$, time-reversal symmetric topological insulators have $\theta = \pi$, and Dirac Semimetals have $\theta = \pi$ (refer to Table~\ref{table1} for $\theta$ values across different materials) \cite{theta1, theta2, theta3, theta4}. In contrast, TWS discussed in this work have a $\theta$ value of $\theta(x, t) = 2 \vec{\mathtt{b}} \cdot \vec{x} - 2\mathtt{b}_0 t$.
\begin{table*}[ht]
\centering 
\begin{tabular}{| c | c |} 
\hline 
$\textcolor{red}{\bf \theta}$ & $\textcolor{red}{\bf Material~Type}$ \\ [0.5ex] 
\hline \hline & \\
$0$ & ~~~~$Ordinary~Insulators$~~~\\
& \\
\hline &\\ 
$\pi$ & ~~~~$Time~Reversal~Symmetric~Topological~ Insulators$~~~\\
& \\
\hline &\\ 
~~$2\mathtt{b}_{\mu}x^{\mu}$~~ & ~~~~$Weyl~Semimetals$~~~\\
& \\
\hline &\\ 
$\pi$ & ~~~~$Dirac~Semimetals$~~~\\
& \\
\hline 
\end{tabular}
\caption{$\theta$-parameters for various material types. These parameters are functions of space and time. Here $\mathtt{b}^{\mu} = (\mathtt{b}_0, \vec{\mathtt{b}})$ and $x^{\mu} = (t, \vec{x})$. Minkowski metric is assumed to take the standart value $\eta=\textrm{diag} (-1, +1, +1, +1)$.} 
\label{table1} 
\end{table*}

\begin{figure}
\centering
\includegraphics[width=6cm]{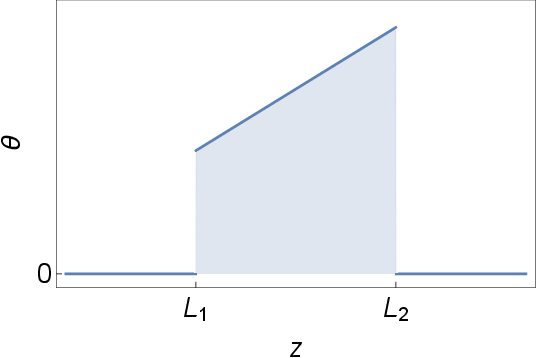}
    \caption{Figure displays the $\theta$-terms corresponding to each specific medium. Colored region specifies TWS. Notice that $\theta$-term is linearly proportional to $2 \mathtt{b}$, where $\mathtt{b}$ implies the distance between Weyl nodes.}
    \label{fig2}
\end{figure}

The TWS slab is constrained between $L_1$ and $L_2$ and  specified by a uniform complex refractive index $\fn$, which is constant throughout the medium. We establish the interaction between the TWS and electromagnetic waves by directing the wave from the left (or right) at an incidence angle, $\phi$, relative to the surface normal of the TWS medium. Electromagnetic wave interactions with this slab are governed by Maxwell's equations, which are modified by topological terms due to magneto-electric optical effects. The topological properties of the slab arise from the arrangement of Weyl nodes, which dictate to appear Fermi arcs on its surfaces. As illustrated in Fig.~\ref{fig1}, these Fermi arcs are found on surfaces aligned with the $z$-axis. Based on our optical design, the Maxwell's equations for this system are formulated as follows\cite{axion3, Oktay2020}:
  \begin{align}
    &\vec{\nabla}\cdot\vec{\cD} = \rho (z) + \beta\,\vec{\mathtt{b}}\cdot\vec{\cB}, &&
    \vec{\nabla}\cdot\vec{\cB} = 0,
    \label{equation1}\\
    &\vec{\nabla} \times \vec{\cH}-\partial_t \vec{\cD}=\vec{\cJ}(z) -\beta\,\vec{\mathtt{b}}\times\vec{\cE}, &&
    \partial_t \vec{\cB}+\vec{\nabla} \times \vec{\cE}=\vec 0,
        \label{equation2}
    \end{align}
Here, $\beta := 2\alpha / \pi Z_0$ is a constant, where $\alpha:=e^2 / 2\varepsilon_0 h c$ is the fine-structure constant, $Z_0:=\sqrt{\mu_0/\varepsilon_0}$ is the vacuum impedance, $e$ is the electron charge, and $c:= 1/\sqrt{\varepsilon_0 \mu_0}$ is the speed of light in a vacuum. The vector $\vec{\mathtt{b}}$ represents the separation between two Weyl nodes aligned in the $z$-direction, given explicitly by $\vec{\mathtt{b}} (z)= \mathtt{b}\,\hat{e}_z$ for $L_1<z<L_2$. The electric and magnetic fields, $\vec{\cE}$ and $\vec{\cB}$ , respectively, are related to the displacement field $\vec{\cD}$ and the magnetic field intensity 
$\vec{\cH}$ through the following constitutive relations:
   \begin{align*}
    \vec{\cD} := \tilde{\varepsilon}\, \vec{\cE}, &&\vec{\cB}:=\tilde{\mu}\vec{\cH},
    \end{align*}
In these descriptions, $\tilde{\varepsilon}$ and $\tilde{\mu}$ represent the permittivity and permeability of the medium through which the electromagnetic wave propagates. They are defined as $\tilde{\varepsilon} := \varepsilon_0 \varepsilon$ and $\tilde{\mu} := \mu_0$, where $\varepsilon$ is the relative permittivity and TWS material has ignorable magnetic property. Specifically, we define $\varepsilon (z) := \varepsilon_b + \frac{i \sigma_{yy}}{\varepsilon_0 \omega}$ for $z \in [L_1, L_2]$. Here, $\varepsilon_b$ represents the contribution from bound charge. It is important to note that within the slab, the quantity $\fn^2 := \varepsilon \mu$ represents the squared complex refractive index $\fn$ of the TWS.

The electric current density in Maxwell's equations is given by $\vec{\cJ}(z) := \sigma(z) \vec{\cE}(z)$ at the surfaces $z = L_1$ and $L_2$, where $\sigma(z)$ represents the surface conductivity at these specific locations on the TWS. In a more formal tensorial notation, the surface current $\vec{\cJ}^s$ can be expressed as $\vec{\cJ}^s_{\alpha} = \sigma^s_{\alpha\beta} \vec{E}_{\beta}$, with the surface conductivities $\sigma^s_{\alpha\beta}$ defined as follows \cite{axion3, conductiv1}:
   \begin{align}
   \sigma^s_{yy} &=\frac{e^2 k_c}{3 \pi h \hat{\omega}_c} \left\{1 -i \left[\hat{\omega}_c^2+\ln\left|1-\hat{\omega}_c^2\right|\right]\right\},   \label{conductivity1}\\
   \sigma^s_{yx} &=\frac{e^2 \mathtt{b}}{\pi h} + \frac{\alpha c}{3 \pi v_F} \ln\left|1-\hat{\omega}_c^2\right|.   \label{conductivity2}
   \end{align}
Here, $\hat{\omega}_c := 2\omega_c/\omega$, where $\omega_c := v_F k_c$ with $v_F$ representing the Fermi velocity and $k_c$ the momentum cut-off, and $k$ is constrained to $k \leq k_c$. It is observed that the component $\sigma^s_{yx}$, as shown in equation (\ref{conductivity2}), governs both Kerr and Faraday rotations within and outside the TWS slab. Additionally, it is important to note that free surface charges accompany the surface currents, and these are interconnected through the continuity equation,
    \be
    \vec{\nabla}\cdot\vec{\cJ}^s + \partial_t \rho^s (z) = 0.
    \ee

We highlight that free charges and currents are generated exclusively on the surfaces of the TWS slab where the incident wave interacts, with no Fermi arcs present. Free charges and currents may also be present inside the TWS, but for the purposes of our analysis, we assume they are not present within the material. This results in the TWS slab behaving as a conductor at its surfaces while acting as a (semi)metal within the bulk of the material.

We now examine the TE mode solutions of Maxwell's equations and consider obliquely incident time-harmonic electromagnetic waves, as illustrated in Fig.~\ref{fig1},
    \begin{align}
    \vec E(\vec{r})=\sE (z)e^{ik_{x}x}\hat e_y.
    \label{ez1}
    \end{align}

The Maxwell equations presented in (\ref{equation1}) and (\ref{equation2}) yield the 3-dimensional Helmholtz equation, which describes the TE mode states and the corresponding magnetic field $\vec{H}$ as follows,
   \begin{align}
   &\left[\nabla^{2} +k^2\varepsilon(z)\mu(z)\right]\vec{E} -i\beta k Z_0 \vec{\mathtt{b}}\times\vec{E} = 0,\label{helmholtz1}\\ &\vec{H} = -\frac{i}{k Z_{0}\mu(z)} \vec{\nabla} \times \vec{E}.\label{helmholtz2}
   \end{align}
Note that the last term is the source of Kerr and Faraday rotations both within and outside the TWS material. These equations are actually coupled when twisted waves occur. Considering Kerr and Faraday rotations, along with the formal similarity between the Helmholtz and Schrödinger equations, it follows directly that Eqs. (\ref{helmholtz1}) and (\ref{helmholtz2}) lead to the following uncoupled equations,
    \be
    -\psi''_{\pm} +  v_{\pm}(\mathbf{z}) \psi_{\pm} = \fK^2 \psi_{\pm}, \label{schro}
    \ee
for the potentials expressed as $v_{\pm}(\mathbf{z}) = \fK^2 \fz_{\pm} (\mathbf{z})$, where $\fz_{\pm} (\mathbf{z})$ is defined as
    \begin{align}
    \fz_{\pm}(\mathbf{z})&:=\left\{\begin{array}{cc}
   \tilde{\fn}_{\pm}^2 & {\rm for}~~~~ \mathbf{z} \in [\mathbf{z}_1, \mathbf{z}_2],\\
    1 & {\rm otherwise.}
    \end{array}\right.\notag
    \end{align}
Here the quantity $\tilde{\fn}_{\pm}$ denotes the effective refractive indices within the material and indicates the birefringence effect, defined as follows
\be
\tilde{\fn}_{\pm} := \sqrt{\tilde{\fn}^2 \pm 2\alpha \mathtt{b} L /\pi \fK \cos\phi}, \qquad  \tilde{\fn} := \sec\phi\sqrt{\fn^2 -\sin^2\phi}. \label{effectiveref}
\ee
In these equations, we use the following scaled variables to simplify the subsequent expressions,
\begin{align}
    &\mathbf{x}:=\frac{x}{L}, &&\mathbf{z}:=\frac{z}{L},   &&\fK:=Lk_z=kL\cos\phi, \label{scaled-var}
    \end{align}
with the identifications $\mathbf{z}_1 := L_1 / L$, $\mathbf{z}_2 := L_2 / L$ and $L:= L_2 - L_1$ representing the length of the slab. We note that the solutions to the Schrödinger equation, $\psi_{\pm}$, correspond to two distinct sets of solutions, referred to as the plus and minus modes, respectively. Therefore, the electric field $\vec{E}$ and magnetic field $\vec{H}$ are computed in their respective components as shown in Table~\ref{table01}.
\begin{table*}[ht]
\centering
{%
\begin{tabular}{@{\extracolsep{4pt}}llllcccc}
\toprule
{} & \multicolumn{1}{c}{Components of $\vec{E}$-field} & {} & \multicolumn{1}{c}{Components of $\vec{H}$-field} & {} \\
 \cline{2-2}
 \cline{3-5}
 \cline{6-8}
   \hline
  & $E_x = \frac{(\mathcal{F}_+ + \mathcal{G}_+)}{2}\,e^{i\fK \mathbf{x}\tan\phi}$ & {} & $H_x = \frac{i\cos\phi}{2 Z_0}\left[\sqrt{\fz_{+}}\mathcal{F}_- - \sqrt{\fz_{-}}\mathcal{G}_-\right]e^{i\fK \mathbf{x}\tan\phi}$ & {} \\
  & $E_y = \frac{-i(\mathcal{F}_+ - \mathcal{G}_+)}{2}\,e^{i\fK \mathbf{x}\tan\phi}$ & {} & $H_y = \frac{\cos\phi}{2 Z_0 \mu}\left[\sqrt{\fz_{+}}\mathcal{F}_- + \sqrt{\fz_{-}}\mathcal{G}_-\right]e^{i\fK \mathbf{x}\tan\phi}$ & {} \\
  & $E_z = 0$ & {} & $H_z = -\frac{i\sin\phi}{2 Z_0}\left[\mathcal{F}_+ - \mathcal{G}_+\right]e^{i\fK \mathbf{x}\tan\phi}$ & {} \\
 \hline
\end{tabular}%
}
\caption{Components of $\vec{E}$ and $\vec{H}$ fields existing inside and outside the TWS slab.}\label{table01}
\end{table*}
Here we introduce the quantities $\mathcal{F}_{\pm}$ and $\mathcal{G}_{\pm}$ in various regions of the optical TWS slab system as follows:
   \begin{flalign}
    \mathcal{F}_{\pm} :=\left\{\begin{array}{ccc}
    A_1^{(+)}\,e^{i\fK\mathbf{z}} \pm C_1^{(+)}\,e^{-i\fK\mathbf{z}} & {\rm for}~~ \mathbf{z} < \mathbf{z}_1,\\
    B_1^{(+)}\,e^{i\fK_+\mathbf{z}} \pm B_2^{(+)}\,e^{-i\fK_+\mathbf{z}} & {\rm for}~~ \mathbf{z}_1 < \mathbf{z} < \mathbf{z}_2 ,\\
    A_2^{(+)}\,e^{i\fK\mathbf{z}} \pm C_2^{(+)}\,e^{-i\fK\mathbf{z}} & {\rm for}~~ \mathbf{z} > \mathbf{z}_2.
    \end{array}\right.
    \qquad
    \mathcal{G}_{\pm}:=\left\{\begin{array}{ccc}
    A_1^{(-)}\,e^{i\fK\mathbf{z}} \pm C_1^{(-)}\,e^{-i\fK\mathbf{z}} & {\rm for}~~ \mathbf{z} < \mathbf{z}_1,\\
    B_1^{(-)}\,e^{i\fK_-\mathbf{z}} \pm B_2^{(-)}\,e^{-i\fK_-\mathbf{z}} & {\rm for}~~ \mathbf{z}_1 < \mathbf{z} < \mathbf{z}_2 ,\\
    A_2^{(-)}\,e^{i\fK\mathbf{z}} \pm C_2^{(-)}\,e^{-i\fK\mathbf{z}} & {\rm for}~~ \mathbf{z} > \mathbf{z}_2.
    \end{array}\right.\notag
    \end{flalign}
Here we specify the quantity $\fK_j$ as given below:
    \be
    \fK_j := \fK \tilde{\fn}_j. \label{ktildej}
    \ee
The complex coefficients $A_j^{(\pm)}, B_j^{(\pm)}$ and $C_j^{(\pm)}$, which depend on $\fK$, are related through the boundary conditions applicable to our TWS slab system. For a detailed account of these boundary conditions, see the Appendix. Accordingly, the transfer matrix can be formulated as follows:
   \be
   \left(
           \begin{array}{c}
             \bA_2 \\
             \bC_2 \\
           \end{array}
   \right) = \mathbb{M} (\fK)\left(
           \begin{array}{c}
             \bA_1 \\
             \bC_1 \\
           \end{array}
   \right), \notag
   \ee
where  $\bA_{j}$ and $\bC_{j}$, with $j=1, 2$, are column matrices representing the coefficients of right- and left-moving waves outside the TWS slab. They are defined as follows:
\begin{align}
&\bA_j = \left(
                       \begin{array}{c}
                         A_j^{(+)} \\
                         A_j^{(-)} \\
                       \end{array}
                     \right), ~~~~~\bC_j = \left(
                       \begin{array}{c}
                         C_j^{(+)} \\
                         C_j^{(-)} \\
                       \end{array}
                     \right),\notag
\end{align}
and $\mathbb{M} (\fK)$ is the $4\times4$ transfer matrix~\cite{prl-2009}, expressed in terms of $2\times2$ matrices for reflection and transmission amplitudes \cite{prl-2009},
\be
\mathbb{M} (\fK) = \left(
                       \begin{array}{cc}
                         \bT^{r}- \bR^{l}\bT^{-l}\bR^{r}& \bR^{l}\bT^{-l} \\
                         -\bT^{-l}\bR^{r} & \bT^{-l} \\
                       \end{array}
                     \right).\label{transmatr}
\ee
The matrices $\bR^{l/r}$ and $\bT^{l/r}$ represent the left/right reflection and transmission amplitudes, respectively, with $2\times 2$ matrix values. We will assume $\bT = \bT^{l} = \bT^{r}$ by applying the reciprocal property of the material.

It’s apparent that we derived the transfer matrix by utilizing standard boundary conditions. Notably, this approach incorporates all parameters of the TWS system within the transfer matrix. By effectively managing these parameters, we can manipulate the transfer matrix to achieve our desired outcomes. To enhance the practicality of these techniques, we will decompose the transfer matrix, revealing a particularly useful decomposition,
\be
\mathbb{M} = \mathcal{U}_2^{-1} \mathcal{U}_1,
\ee
where 
\be
 \mathcal{U}_j := \mathbb{P}_{j\backslash}\,\mathbb{S}_1 + \mathbb{P}_{j\slash}\,\mathbb{S}_2.
\ee
Here, decomposition matrices $\mathbb{P}_{j\backslash}$, $\mathbb{P}_{j\slash}$, $\mathbb{S}_1$ and $\mathbb{S}_2$ are represented as follows:
\begin{align}
 \mathbb{P}_{1\backslash} &:=
    \begin{pmatrix}
    e^{i \fK_{ab+}^{(+)}} & 0 & 0 & 0 \\
    0 & e^{i \fK_{ab+}^{(-)}} & 0 & 0 \\
    0 & 0 & e^{-i \fK_{ab-}^{(-)}} & 0 \\
    0 & 0 & 0 & e^{-i \fK_{ab-}^{(+)}}
    \end{pmatrix},\qquad  &&\mathbb{P}_{1\slash} :=
    \begin{pmatrix}
    e^{-i \fK_{ab+}^{(-)}} & 0 & 0 & 0 \\
    0 & e^{-i \fK_{ab-}^{(+)}} & 0 & 0 \\
    0 & 0 & e^{i \fK_{ab-}^{(+)}} & 0 \\
    0 & 0 & 0 & e^{i \fK_{ab-}^{(-)}}
    \end{pmatrix}, \notag\\
  \mathbb{P}_{2\backslash} &:=
    \begin{pmatrix}
    e^{i \fK_{ba+}^{(+)}} & 0 & 0 & 0 \\
    0 & e^{i \fK_{ba+}^{(-)}} & 0 & 0 \\
    0 & 0 & e^{-i \fK_{ba-}^{(-)}} & 0 \\
    0 & 0 & 0 & e^{-i \fK_{ba-}^{(+)}}
    \end{pmatrix},\qquad  &&\mathbb{P}_{2\slash} :=
    \begin{pmatrix}
    e^{-i \fK_{ba+}^{(-)}} & 0 & 0 & 0 \\
    0 & e^{-i \fK_{ba-}^{(+)}} & 0 & 0 \\
    0 & 0 & e^{i \fK_{ba-}^{(+)}} & 0 \\
    0 & 0 & 0 & e^{i \fK_{ba-}^{(-)}} 
    \end{pmatrix}, \notag\\  
  \mathbb{S}_{1} &:=
    \begin{pmatrix}
    \tilde{\fn}_+ + \mu (1 + \sigma_+) & -\mu \sigma_- & 0 & 0 \\
    \tilde{\fn}_+ - \mu (1 + \sigma_+) & \mu \sigma_- & 0 & 0 \\
    0 & 0 & -\mu \sigma_+ & \tilde{\fn}_- - \mu (1 - \sigma_-) \\
    0 & 0 & \mu \sigma_+ & \tilde{\fn}_- + \mu (1 - \sigma_-)
    \end{pmatrix},\qquad  &&\mathbb{S}_{2} :=
    \begin{pmatrix}
    0 & 0 & \tilde{\fn}_+ - \mu (1 - \sigma_+) & -\mu \sigma_- \\
    0 & 0 & \tilde{\fn}_+ + \mu (1 - \sigma_+) & \mu \sigma_- \\
    -\mu \sigma_+ & \tilde{\fn}_- + \mu (1 + \sigma_-) & 0 & 0 \\
    \mu \sigma_+ & \tilde{\fn}_- - \mu (1 + \sigma_-) & 0 & 0
    \end{pmatrix}. \notag
\end{align}
In these expressions, we define $\fK_{x y \ell}^{(j)} := \fK (x + j y \tilde{\fn}_{\ell})$, where $x$ and $y$ belong to $\{a, b\}$, and $j$ and $\ell$ take values in $\{ +, -\}$. Notice that we employ the identifications $L_1=a$ and $L_2 = b$ for convenience\footnote{Note the abuse of notation: while $\mathtt{b}$ is typically used to denote the distance between Weyl nodes, here $b$ represents the value of the right boundary of the TWS slab.}. Thus, It turns out that the matrix $\mathbb{M}$ can be expressed in the following decomposition:
\begin{align}
\mathbb{M} &= \left(\mathbb{S}_{1}^{-1} \mathbb{P}_{2\backslash}^{-1} + \mathbb{S}_{2}^{-1} \mathbb{P}_{2\slash}^{-1} \right) \left( \mathbb{P}_{1\backslash} \mathbb{S}_{1} + \mathbb{P}_{1\slash} \mathbb{S}_{2}  \right), \notag\\
&= \mathbb{S}_{1}^{-1} \mathbb{P}_{2\backslash}^{-1} \mathbb{P}_{1\backslash} \mathbb{S}_{1} + \mathbb{S}_{1}^{-1} \mathbb{P}_{2\backslash}^{-1} \mathbb{P}_{1\slash} \mathbb{S}_{2} + \mathbb{S}_{2}^{-1} \mathbb{P}_{2\slash}^{-1} \mathbb{P}_{1\backslash} \mathbb{S}_{1} + \mathbb{S}_{2}^{-1} \mathbb{P}_{2\slash}^{-1} \mathbb{P}_{1\slash} \mathbb{S}_{2}.
\end{align}
The generalized transfer matrix $\mathbb{M}$ for the whole system of a single TWS slab consists of the sum of four unique similarity transformations, as demonstrated. Every term in this expression serves a specific purpose, with each one carrying out a different similarity transformation.

Having established the transfer matrix for a single TWS slab, we can now use it to derive the transfer matrix for multiple slabs arranged side by side. This allows us to leverage the beneficial associative property of the transfer matrix. We will explore this analysis in detail in the following section.

\section{$\mathcal{PT}$-Symmetric TWS}
\label{S22}

\begin{figure}
    \begin{center}
    \includegraphics[scale=.255]{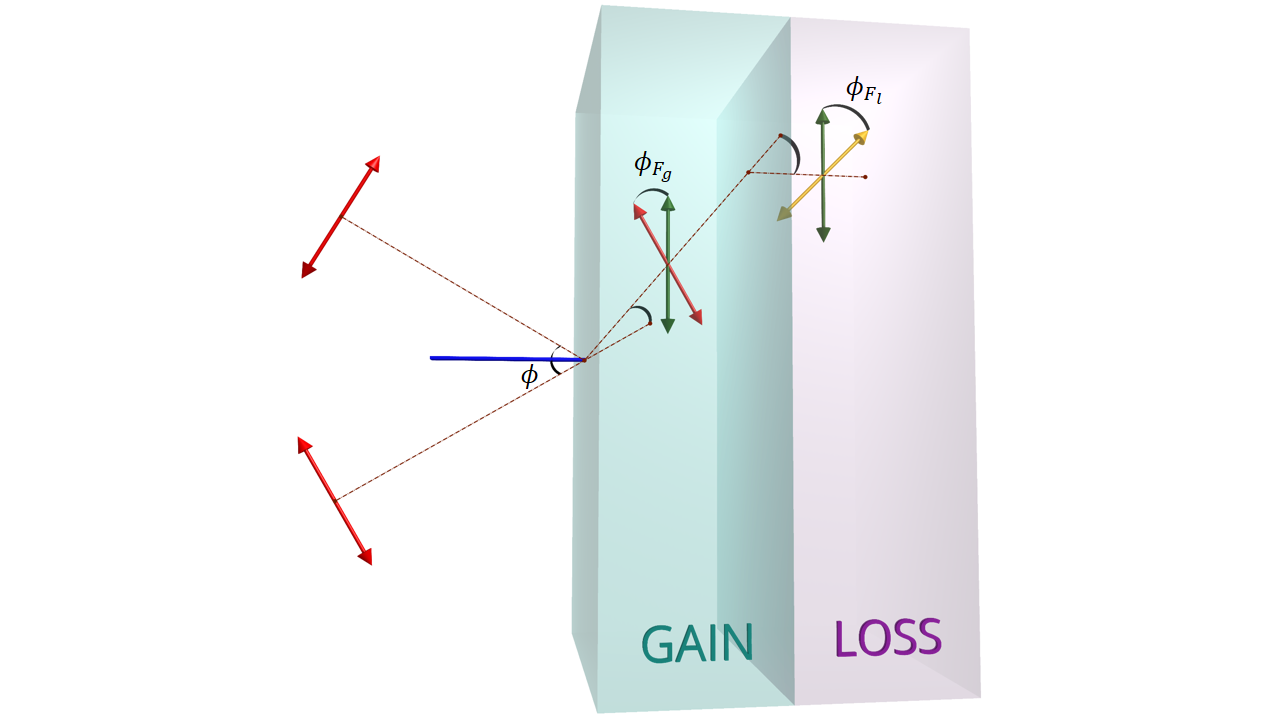}
    \caption{(Color online) The TE mode configuration for the interaction of an electromagnetic wave with the Weyl semimetal slab (Left panel). The wave is emitted on the slab by an angle $\phi$ which is measured from the normal to the surface, and direction of the polarization is rotated by an angle of $\phi_F$ and $\phi_K$ inside and outside of the slab respectively. On the right panel, Fermi arcs due to Weyl nodes for an individual slab are presented.}
    \label{figpt}
    \end{center}
    \end{figure}

To construct a $\mathcal{PT}$-symmetric TWS system, it is essential to arrange the gain and loss components side by side, ensuring they are equal in strength. This configuration not only satisfies the criteria for $\mathcal{PT}$ symmetry but also highlights the importance of adhering to the $V (\mathbf{z}) = V^*(-\mathbf{z})$ condition for the potential of the system. Understanding these foundational elements is crucial for achieving the desired symmetry in the system's behavior. Figure~(\ref{figpt}) displays our $\mathcal{PT}$-symmetric TWS system, featuring equal amounts of gain and loss components. To better understand the non-Hermitian nature of the system's scattering states, it is necessary to construct the total transfer matrix. Denoted as $\mathbb{M}_{\mathcal{PT}}$, this total transfer matrix can be expressed in terms of the individual transfer matrices as follows,
\begin{align}
\mathbb{M}_{\mathcal{PT}} = \mathbb{M}_{loss}\,\mathbb{M}_{gain}.\label{tmt}
\end{align}
This utilizes the associative property of the transfer matrix, which is particularly advantageous because it allows us to easily combine complex systems into a single, unified system. In our study, we set $a = 0$ and $b = 1$, and use the refractive index $\fn_{gain} = \fn$ for the gain component. For the loss component, we take $b= 1$ and $c = 2$ as the boundary values, while evaluating the refractive index as $\fn_{loss} = \fn^*$ \footnote{Keep in mind that the scaled variables $a = 0$, $b = 1$, and $c = 2$ correspond to the original values of $0$, $L$, and $2L$, respectively.}.

Our goal now is to ensure that the TWS system delivers results in the intended direction by precisely controlling the transfer matrix. Since the $\mathcal{PT}$-symmetric TWS system is a non-Hermitian physical scenario, examining its exceptional points becomes essential.

The transfer matrix serves as a vital tool, encompassing all the fundamental information related to a scattering system. This information can be utilized to manipulate the system's behavior following scattering, as all the system's parameters are contained within the transfer matrix. Thus, the exceptional points of the $\mathcal{PT}$-symmetric TWS environment in the context of non-Hermitian physics can be rederived through the transfer matrix.

The transfer matrix described in equation (\ref{tmt}) contains comprehensive information about the $\mathcal{PT}$-symmetric TWS scattering system. This section explores how to identify spectral singularities using the transfer matrix. In gain-doped TWS systems, spectral singularities are linked to exceptional points or non-Hermitian phases, where the system exhibits a typical lasing threshold behavior, corresponding to zero-width resonances at real energy eigenstates\footnote{Although this study considers the concept of a topological laser primarily in terms of the gain value around the lasing threshold condition, a more comprehensive understanding of the concept requires evaluating additional parameters such as the threshold power and the single-mode radiation spectrum. Addressing these aspects in a different context would better delineate the scope and limitations of the present work.}. Our analysis reveals that the one-dimensional TWS system effectively becomes two-dimensional due to Faraday rotation, even when electromagnetic waves are incident in the TE mode. As a result, the transfer matrix takes on a $4\times 4$ matrix form. Exceptional points can be detected by imposing constraints on the components of the transfer matrix. Exceptional points are characterized by state vectors that form the Hilbert space of the system. In a standard Hermitian system, these state vectors, which should be orthonormal, become parallel and/or have coincident eigenvalues, leading to exceptional points. In our system, four orthonormal state vectors arise from two distinct modes, allowing for a variety of exceptional point configurations. This study focuses specifically on a case known as spectral singularity. After scattering, spectral singularities in the system manifest as outwardly propagating wave configurations. By distinguishing between the different solution modes—termed plus (+) and minus (-) modes-varied spectral singularity scenarios can be realized. By “plus mode,” we refer to the wave modes generated by the waves resulting from the refractive index $\tilde{\fn}_{+}$. Similarly, by “minus mode,” we mean the wave modes arising due to the refractive index $\tilde{\fn}_{-}$. The significance of these modes lies in the fact that their laser output characteristics differ substantially, owing to the distinct behaviors of the respective refractive indices. These unique cases will be examined in detail in the following sections, along with practical examples from $\mathcal{PT}$-symmetric TWS systems.

We will first define the gain coefficient of the TWS system in the following manner\cite{Oktay2020, CPA, silfvast}:
\begin{equation}
g := -2k\kappa ,
\end{equation}
where $\kappa$ represents the imaginary part of the refractive index $ \mathbf{n}$, defined as $\mathbf{n} = \eta + i \kappa$. To examine how the physical parameters of our system affect the identified spectral singularity condition, we will consider the TaAs material within a slab environment as an example. TaAs is notable for being one of the first practical topological Weyl semimetals \cite{taas1, topweyl12, taas3, taas4, taas5, taas6}. Its characteristics are detailed as follows \cite{buckeridge, ramshaw, dadsetani, silfvast, kriegel}:
\begin{align} 
\eta&=6,~~ L=500~\textrm{nm},~~ \phi=30^{\circ}, \label{specification}
\end{align}
We observe that the refractive index of the TWS varies with the Fermi energy, ambient temperature, and light frequency \cite{taasp1, taasp2, taasp3}. Additionally, we need the conductivity of TaAs for our computational analysis, as outlined in Eqns.~(\ref{conductivity1}) and (\ref{conductivity2}).

\section{Various Spectral Singularity Configurations: Types of Topological Lasers}

\subsection{A. Plus Mode Spectral Singularity Configuration: Plus-Mode Topological TWS Laser}

\begin{figure*}
\centering
\includegraphics[width=5cm]{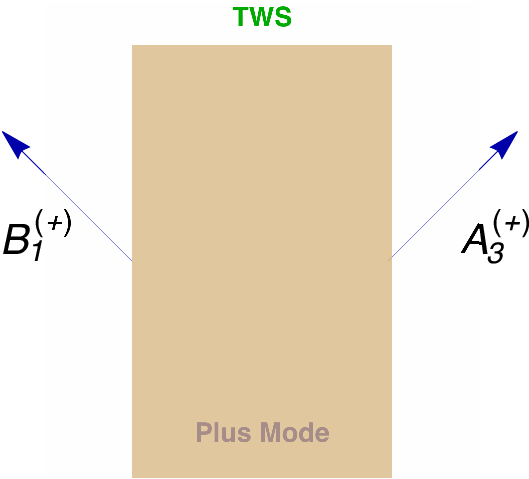}~~
\includegraphics[width=5cm]{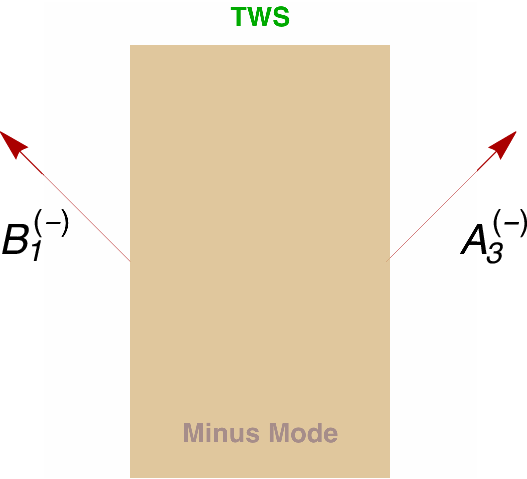}~
\includegraphics[width=5cm]{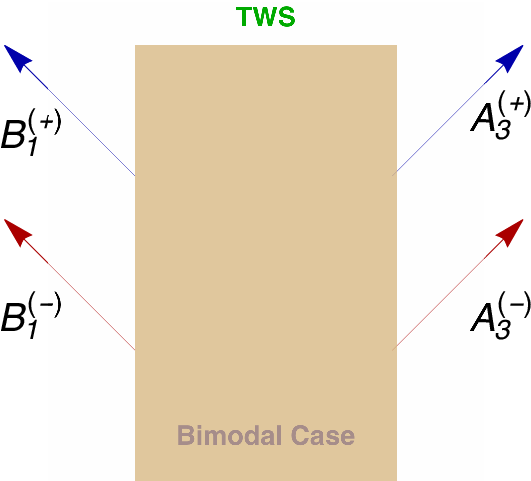}
    \caption{(Color Online) The figure shows the spectral singularity configurations and laser output modes of the Plus Mode (left panel), Minus Mode (middle panel) and Bimodal case (right panel) in the TWS medium. Waves with amplitudes $B_{1}^{(+)}$ and $A_{3}^{(+)}$ are output from the left and right sides in the Plus Mode, amplitudes $B_{1}^{(-)}$ and $A_{3}^{(-)}$ for the Minus Mode, and amplitudes $B_{1}^{(+)}$, $B_{1}^{(-)}$, $A_{3}^{(+)}$, and $A_{3}^{(-)}$ for the Bimodal case,  respectively.}
    \label{fig41}
\end{figure*}

In this case, there are only the outgoing waves of the Plus Mode on the far right and far left sides of the system (\ref{fig41}). Thus, only the amplitudes $C_{1}^{(+)}$ and $A_{3}^{(+)}$ of waves remain. For this to happen, $A_{1}^{(+)} = A_{1}^{(-)} = C_{1}^{(-)} = C_{3}^{(+)} = A_{3}^{(-)} = C_{3}^{(-)} = 0$ must be present. This can only happen if
\begin{equation}
\mathbb{M}_{23} = \mathbb{M}_{33} = \mathbb{M}_{43} = 0,
\end{equation}
provided. Here the symbols $\mathbb{M}_{ij}$ denotes the $i, j$ components of the matrix $\mathbb{M}$. These conditions are the Plus Mode spectral singularity conditions.

\subsection{B. Minus Mode Spectral Singularity Configuration: Minus-Mode Topological TWS Laser}

In this case, the only waves present are outgoing Minus-Mode waves located at both the far left and far right edges of the system, as shown in Fig~.\ref{fig41}. These waves have amplitudes $C_{1}^{(-)}$ and $A_{3}^{(-)}$. For this to occur, the conditions $A_{1}^{(+)} = A_{1}^{(-)} = C_{1}^{(+)} = C_{3}^{(+)} = A_{3}^{(+)} = C_{3}^{(-)} = 0$ must be satisfied. This can only happen if
\begin{equation}
\mathbb{M}_{14} = \mathbb{M}_{34} = \mathbb{M}_{44} = 0,
\end{equation}
is provided for real $k$ values. These conditions are the spectral singularity conditions for the Minus Mode.

\subsection{C. Bimodal Spectral Singularity Configuration: Bimodal Topological TWS Laser}

In this case, the system has only outgoing waves of the bimodal case on the far right and far left sides, as shown in Fig.~\ref{fig41}. These waves have amplitudes $C_{1}^{(+)}$, $C_{1}^{(-)}$, $A_{3}^{(+)}$, and $A_{3}^{(-)}$. For this to occur, the conditions $A_{1}^{(+)} = A_{1}^{(-)} = C_{3}^{(+)} = C_{3}^{(-)} = 0$ must be satisfied. This can only happen if
\begin{equation}
\mathbb{M}_{33} = \mathbb{M}_{34} = \mathbb{M}_{43} = \mathbb{M}_{44} = 0,
\end{equation}
is provided for real $k$-values. These conditions are the spectral singularity conditions for the  bimodal configuration.

\subsection{D. Random Spectral Singularity Configurations: Random Topological TWS Lasers} 

The setup we've analyzed enables us to create random spectral singularities, making it possible to generate random laser beam outputs from either side of the TWS,  irrespective of any particular lasing mode. For instance, one could create a topological system that emits laser light exclusively from either the right or left side. This approach illustrates the effectiveness of a topological laser configuration for achieving the desired outcomes and demonstrates the potential of our method to yield highly diverse results. As shown in Table~(\ref{table3}), 15 different lasing conditions can be achieved using a TWS material. Some of these conditions are appropriate for unidirectional lasing, while others are suitable for bidirectional lasing or random lasing. Achieving a specific lasing configuration requires meeting the relevant spectral singularity condition. However, as can be seen from the Table~(\ref{table3}), it is not possible to obtain laser states that only exit from the right side for the wave configurations emitted from the left side. Similarly, for the waves emitted from the right side, only laser configurations that emerge from the left side are not permitted. In this case, the 15 distinct topological laser states observed are actually reduced to 12.

\begin{table}[ht]
\centering 
\begin{tabular}{| c | c | c | c |} 
\hline 
$\textcolor{blue}{\bf Type~of~Laser}$ & $\textcolor{blue}{\bf Left~Side}$ & $\textcolor{blue}{\bf Right~Side}$ & $\textcolor{blue}{\bf Spectral~Singularity~Condition}$\\ [0.5ex] 
\hline \hline \cellcolor{type1} & & &\\
\cellcolor{type1}$Unidirectional~Laser~from~Left~(+~Mode)$ & ~~~~$+$~~~ & ~~~~$None$~~~ & $\mathbb{M}_{13} = \mathbb{M}_{23} = \mathbb{M}_{33} = \mathbb{M}_{43} = 0$ \\
\cellcolor{type1} &  & &\\
\hline \cellcolor{type2} & & &\\ 
\cellcolor{type2}$Unidirectional~Laser~from~Left~(-~Mode)$ & ~~~~$-$~~~ & ~~~~$None$~~~ & $\mathbb{M}_{14} = \mathbb{M}_{24} = \mathbb{M}_{34} = \mathbb{M}_{44} = 0$\\ \cellcolor{type2}& & &\\
\hline \cellcolor{type3} & & & $\mathbb{M}_{13} = \mathbb{M}_{23} = \mathbb{M}_{33} = \mathbb{M}_{43} = 0$\\ 
\cellcolor{type3} $Unidirectional~Laser~from~Left~(Bimodal)$ & ~~~~$+~\&~-$~~~ & ~~~~$None$~~~ & \\
\cellcolor{type3} & & & $\mathbb{M}_{14} = \mathbb{M}_{24} = \mathbb{M}_{34} = \mathbb{M}_{44} = 0$\\
\hline & & &\\ 
$Unidirectional~Laser~from~Right~(+~Mode)$ & ~~~~$None$~~~ & ~~~~$+$~~~ & $\textcolor{red} {\textrm{NOT~allowed}}$\\
& & &\\
\hline & & &\\
$Unidirectional~Laser~from~Right~(-~Mode)$ & ~~~~$None$~~~ & ~~~~$-$~~~ & $\textcolor{red} {\textrm{NOT~allowed}}$\\
& & &\\
\hline & & &\\
$Unidirectional~Laser~from~Right~(Bimodal)$ & ~~~~$None$~~~ & ~~~~$+~\&~-$~~~ & $\textcolor{red} {\textrm{NOT~allowed}}$\\
& & &\\
\hline \cellcolor{type4} & & &\\
\cellcolor{type4} $Bidirectional~Laser~(+~Mode)$ & ~~~~$+$~~~ & ~~~~$+$~~~ & $\mathbb{M}_{23} = \mathbb{M}_{33} = \mathbb{M}_{43} = 0$ \\
\cellcolor{type4} & & &\\
\hline \cellcolor{type5} & & &\\
\cellcolor{type5} $Bidirectional~Laser~(+~from~Left,~-~from~Right)$ & ~~~~$+$~~~ & ~~~~$-$~~~ & $\mathbb{M}_{13} = \mathbb{M}_{33} = \mathbb{M}_{43} = 0$\\
\cellcolor{type5} & & &\\
\hline \cellcolor{type6} & & &\\
\cellcolor{type6}  $Bidirectional~Laser~~(-~from~Left,~+~from~Right)$ & ~~~~$-$~~~ & ~~~~$+$~~~ & $\mathbb{M}_{24} = \mathbb{M}_{34} = \mathbb{M}_{44} = 0$\\
\cellcolor{type6}  & & &\\
\hline \cellcolor{type7} & & &\\
\cellcolor{type7} $Bidirectional~Laser~(-~Mode)$ & ~~~~$-$~~~ & ~~~~$-$~~~ & $\mathbb{M}_{14} = \mathbb{M}_{34} = \mathbb{M}_{44} = 0$ \\
\cellcolor{type7} & & &\\
\hline \cellcolor{type8} & & &\\
\cellcolor{type8} $Bidirectional~Laser~(+~from~Left,~+\&-~from~Right)$ & ~~~~$+$~~~ & ~~~~$+~\&~-$~~~ & $\mathbb{M}_{33} = \mathbb{M}_{43} = 0$\\
\cellcolor{type8} & & &\\
\hline \cellcolor{type9} & & &\\
\cellcolor{type9} $Bidirectional~Laser~(-~from~Left,~+\&-~from~Right)$ & ~~~~$-$~~~ & ~~~~$+~\&~-$~~~ & $\mathbb{M}_{34} = \mathbb{M}_{44} = 0$\\
\cellcolor{type9} & & &\\
\hline \cellcolor{type10} & & & $\mathbb{M}_{23} = \mathbb{M}_{24} = 0$\\
\cellcolor{type10} $Bidirectional~Laser~(+\&-~from~Left,~+~from~Right)$ & ~~~~$+~\&~-$~~~ & ~~~~$+$~~~ & $\mathbb{M}_{33} = \mathbb{M}_{34} = 0$\\
\cellcolor{type10} & & & $\mathbb{M}_{43} = \mathbb{M}_{44} = 0$\\
\hline \cellcolor{type11} & & & $\mathbb{M}_{13} = \mathbb{M}_{14} = 0$\\
\cellcolor{type11} $Bidirectional~Laser~(+\&-~from~Left,~-~from~Right)$ & ~~~~$+~\&~-$~~~ & ~~~~$-$~~~ & $\mathbb{M}_{33} = \mathbb{M}_{34} = 0$\\
\cellcolor{type11} & & & $\mathbb{M}_{43} = \mathbb{M}_{44} = 0$\\
\hline \cellcolor{type12}& & &\\
\cellcolor{type12}$Bidirectional~Laser~(Bimodal)$& ~~~~$+~\&~-$~~~ & ~~~~$+~\&~-$~~~ & $\mathbb{M}_{33} = \mathbb{M}_{34} = \mathbb{M}_{43} = \mathbb{M}_{44} = 0$ \\
\cellcolor{type12}& & &\\
\hline 
\end{tabular}
\caption{(Color Online) The table presents all potential laser output configurations and conditions from both sides of the TWS slab for an electromagnetic wave emitted from the left side. Please note that the $\mathbb{M}$ matrix used here actually implies $\mathbb{M}_{\mathcal{PT}}$, i.e. $\mathbb{M}_{\mathcal{PT}} = \mathbb{M}_{loss}\,\mathbb{M}_{gain}$. The colors displayed here correspond to the laser types that signify the reel zeros of the transfer matrix components shown in Fig.~\ref{transfermatr}.} 
\label{table3} 
\end{table}

\begin{figure*}
\centering
\includegraphics[width=6cm]{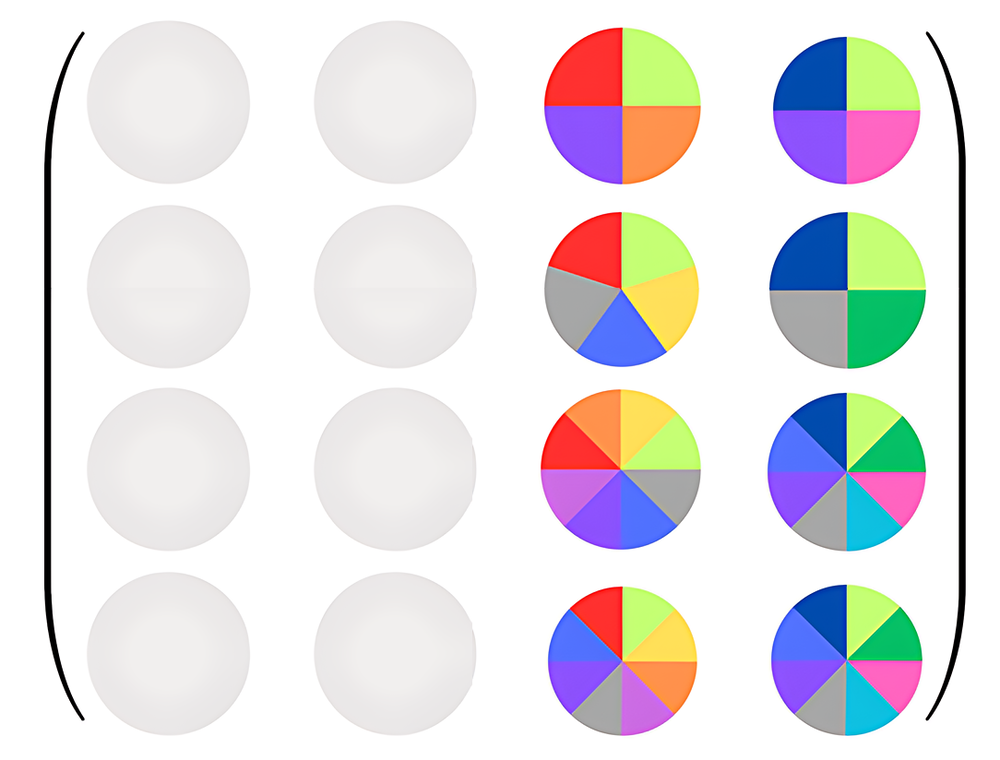}
    \caption{(Color Online) The figure displays a diagram of TWS laser types within the components of the transfer matrix $\mathbb{M}_{\mathcal{PT}}$ for a $\mathcal{PT}$-symmetric TWS environment. It illustrates the laser generation conditions for each TWS laser type listed in Table~(\ref{table3}), with different colors representing different laser types. Notably, only the $3^{rd}$ and $4^{th}$ columns of the transfer matrix generate laser beams. Please note that the colors shown here correspond to the laser types colored differently in Table~\ref{table3}.}
    \label{transfermatr}
\end{figure*}

The figure~(\ref{transfermatr}) reveals that the configurations presented in Table~(\ref{table3}) originate from diverse combinations of the components in the last two columns of the transfer matrix. This highlights that these columns play a pivotal role in determining all the distinct laser configurations in TWS slab. 

The configuration of a $\mathcal{PT}$-symmetric TWS laser, influenced by system parameters, depends on first identifying the laser type and then applying the relevant conditions. Out of the 12 laser types, we will concentrate on two main categories: Plus-Mode and Minus-Mode topological TWS lasers. Similar analyses can also be conducted for other laser types.

\begin{figure*}[ht!] 
    \centering
        \begin{tikzpicture} 
        \node[anchor=north west,inner sep=0pt] at (0,0){\includegraphics[width=5.3cm]{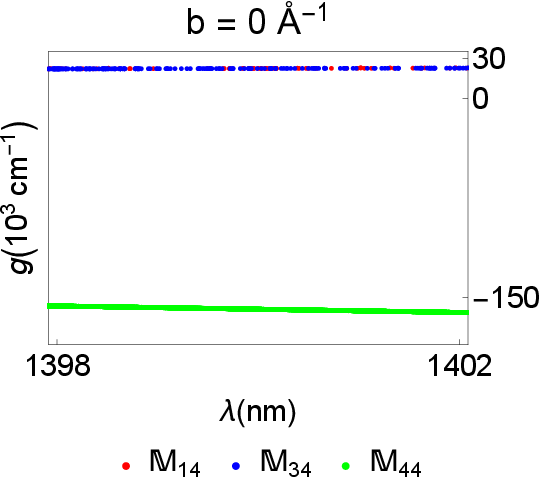}};
        \node[font=\sffamily\bfseries\normalsize] at (6ex, -1.5ex) {(a)}; \draw[orange, thick,->] (2.58,-4.85) -- (2.58,-5.65);
        \end{tikzpicture} 
         \begin{tikzpicture}  
        \node[anchor=north west,inner sep=0pt] at (0,0){\includegraphics[width=5cm]{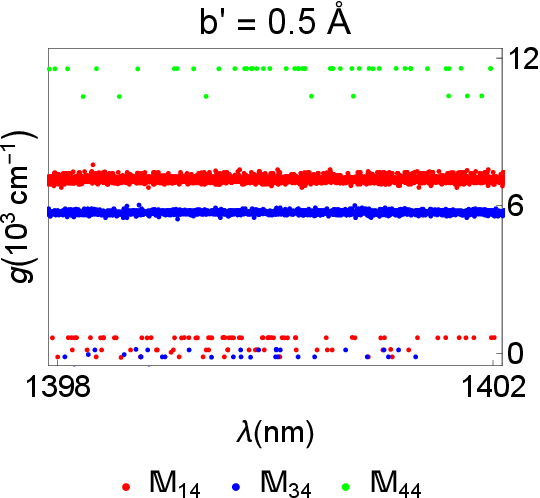}};
        \node[font=\sffamily\bfseries\normalsize] at (6ex, -1.5ex) {(b)}; \draw[orange, thick,->] (2.58,-4.85) -- (2.58,-5.65);
    \end{tikzpicture}
                 \begin{tikzpicture} 
        \node[anchor=north west,inner sep=0pt] at (0,0){\includegraphics[width=5cm]{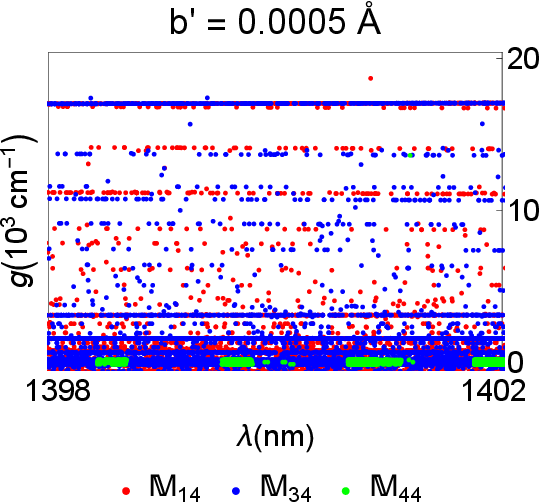}}; 
        \node[font=\sffamily\bfseries\normalsize] at (6ex,-1.5ex) {(c)}; \draw[orange, thick,->] (2.58,-4.85) -- (2.58,-5.65);
    \end{tikzpicture}\\ 
     \begin{tikzpicture} 
        \node[anchor=north west,inner sep=0pt] at (0,0){\includegraphics[width=5cm]{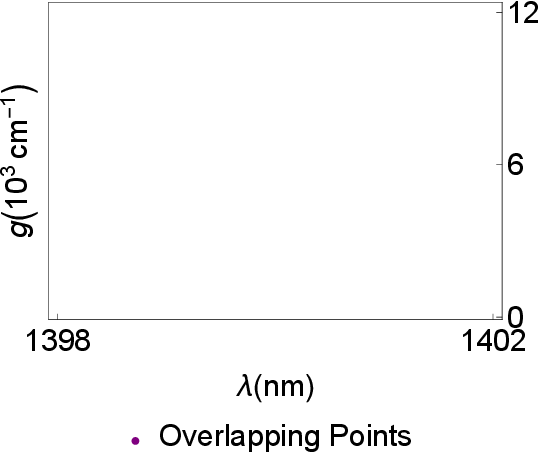}};
        \node[font=\sffamily\bfseries\normalsize] at (6ex,1.5ex) {(d)};
        \end{tikzpicture}
     \begin{tikzpicture} 
        \node[anchor=north west,inner sep=0pt] at (0,0){\includegraphics[width=5cm]{glamdaminusb05intersection.eps}};
        \node[font=\sffamily\bfseries\normalsize] at (6ex,1.5ex) {(e)};
        \end{tikzpicture}\ 
     \begin{tikzpicture} 
        \node[anchor=north west,inner sep=0pt] at (0,0){\includegraphics[width=5cm]{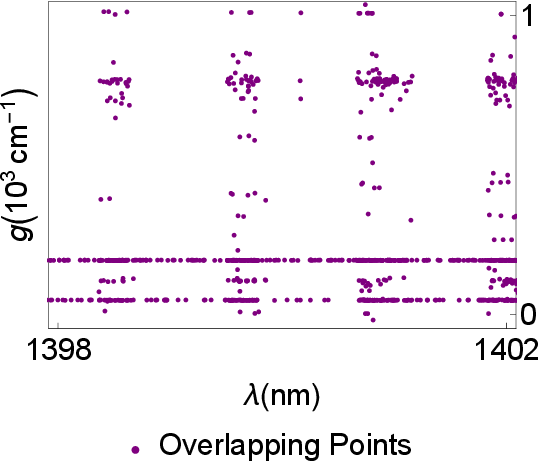}};
        \node[font=\sffamily\bfseries\normalsize] at (6ex,1.5ex) {(f)};
        \end{tikzpicture}
 \caption{(Color Online) In the graphs, spectral singularities are displayed over the $\lambda - g$ plane of the Minus Mode configuration corresponding to distinct $\mathtt{b}'$ values. The graphs in the upper panels correspond to the values $\mathtt{b}'= \infty~{\AA}$ (or $\mathtt{b}= 0~{\AA}^{-1}$), $\mathtt{b}'= 0.5~{\AA}$ and $\mathtt{b}'= 0.0005~{\AA}$ respectively, and each color corresponds to a different real zero component of the transfer matrix. Here, the blue color belongs to the zeros of the $\mathbb{M}_{14}$, the red color to the $\mathbb{M}_{34}$ and the green color to the $\mathbb{M}_{44}$ components. Graphs in the lower panel indicate the intersection points of the points shown in different colors in the upper panels. Notice that graphs are drawn using the data in (\ref{specification}). As can be seen from the lower graphs, only special values of $\mathtt{b}'$ (In our case $\mathtt{b}'= 0.0005~{\AA}$) allow laser beam exit from both sides of the slab. Also note the topological character of the arrangement of spectral singularity points. Please note that the colors presented here are unrelated to those shown in the previous table.}\label{figg1}
\end{figure*}

\begin{figure*}[ht!]
    \centering
        \begin{tikzpicture} 
        \node[anchor=north west,inner sep=0pt] at (0,0){\includegraphics[width=5cm]{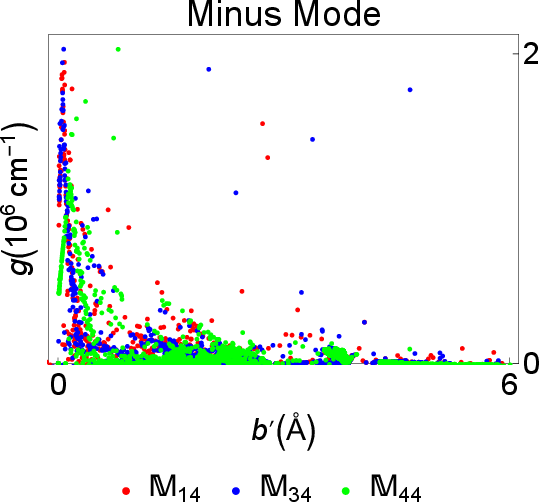}};
        \node[font=\sffamily\bfseries\normalsize] at (6ex, -1.3ex) {(a)}; \draw[orange, thick,->] (5.48,-1.9) -- (6.48,-1.9);
        \end{tikzpicture} ~~~
         \begin{tikzpicture}  
        \node[anchor=north west,inner sep=0pt] at (0,0){\includegraphics[width=5cm]{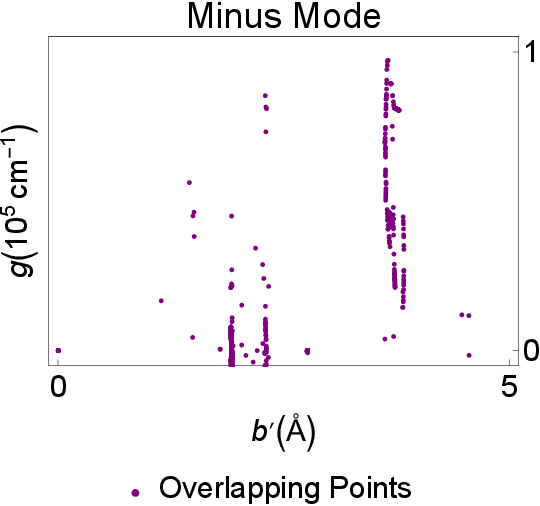}};
        \node[font=\sffamily\bfseries\normalsize] at (6ex, -1.3ex) {(b)}; 
    \end{tikzpicture}
  \caption{(Color Online) In the graphs, spectral singularities are displayed over the $\mathtt{b}' - g$ plane of the Minus Mode configuration corresponding to resonance wavelength $\lambda = 1400~\textrm{nm}$. The graph in the left panels corresponds to a different real zero components of the transfer matrix. Here, the blue color belongs to the zeros of the $\mathbb{M}_{14}$, the red color to the $\mathbb{M}_{34}$ and the green color to the $\mathbb{M}_{44}$ components. Graph in the right panel indicate the intersection points of the points shown in different colors in the left panel. As can be seen from the right graph, only special values of $\mathtt{b}'$ allow laser beam exit from both sides of the slab, verifying the observation found in Fig.~ (\ref{figg1}). Also note the topological character of the arrangement of spectral singularity points.} \label{figg2}
\end{figure*}
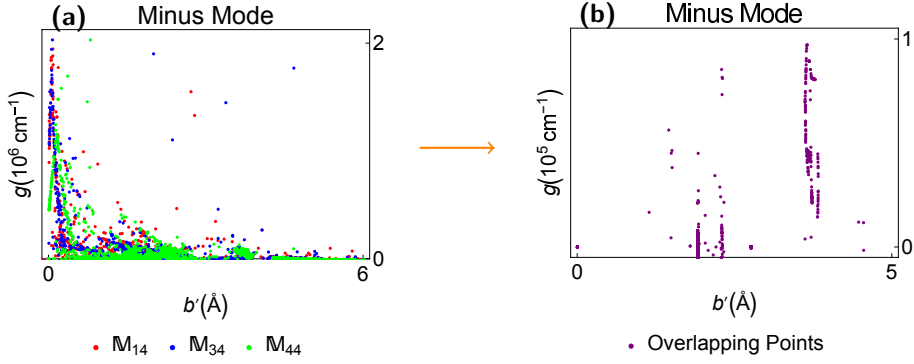

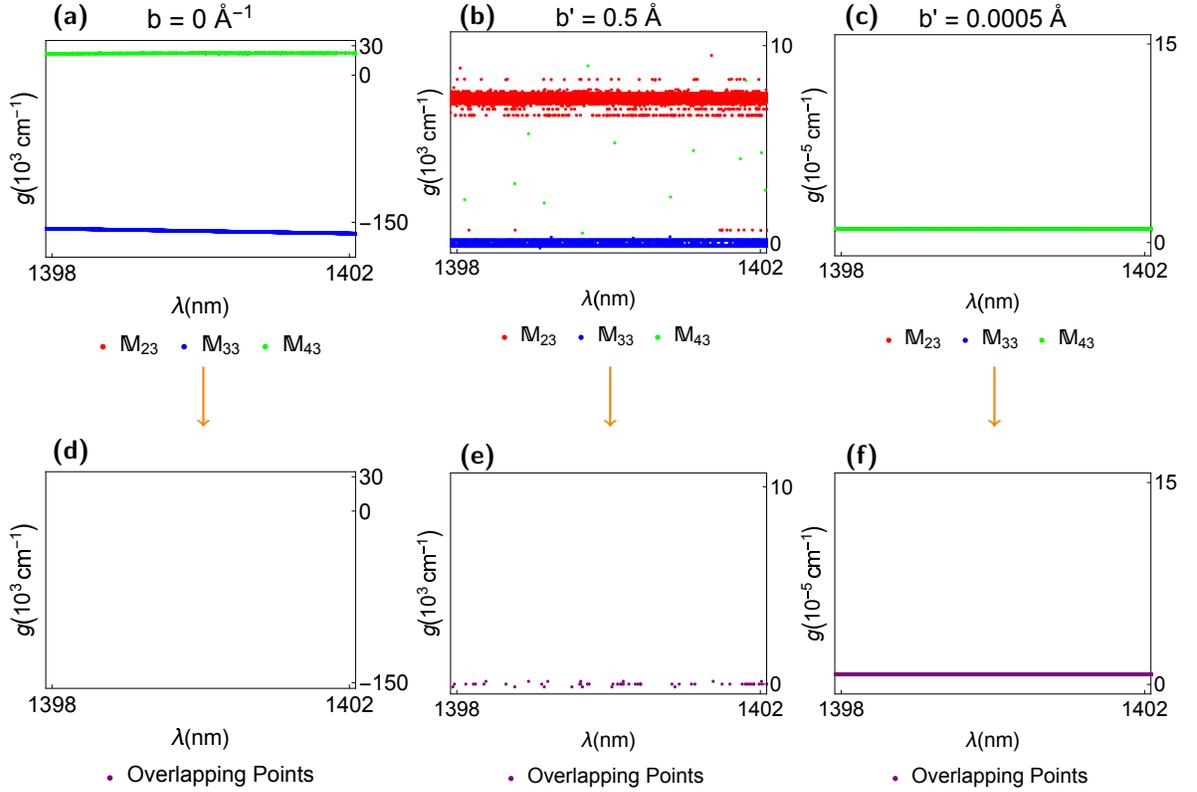
\begin{figure*}[ht!]
    \centering
        \begin{tikzpicture} 
        \node[anchor=north west,inner sep=0pt] at (0,0){\includegraphics[width=5.3cm]{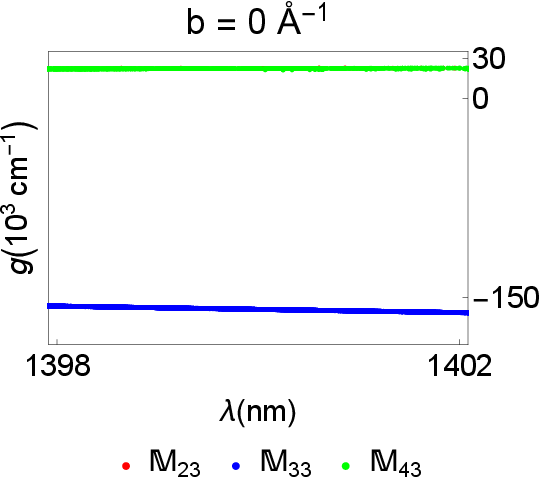}};
        \node[font=\sffamily\bfseries\normalsize] at (6ex, -1.5ex) {(a)}; \draw[orange, thick,->] (2.58,-4.85) -- (2.58,-5.65);
        \end{tikzpicture} 
         \begin{tikzpicture}  
        \node[anchor=north west,inner sep=0pt] at (0,0){\includegraphics[width=5cm]{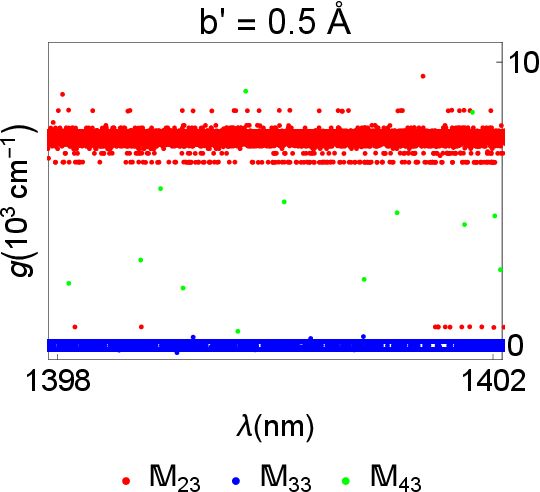}};
        \node[font=\sffamily\bfseries\normalsize] at (6ex, -1.5ex) {(b)}; \draw[orange, thick,->] (2.58,-4.85) -- (2.58,-5.65);
    \end{tikzpicture}
                 \begin{tikzpicture} 
        \node[anchor=north west,inner sep=0pt] at (0,0){\includegraphics[width=5cm]{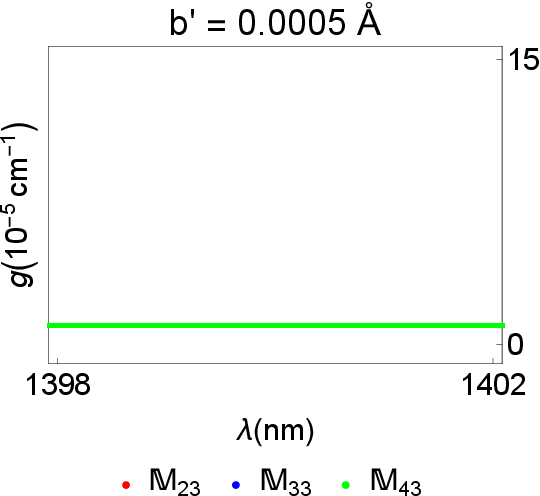}}; 
        \node[font=\sffamily\bfseries\normalsize] at (6ex,-1.5ex) {(c)}; \draw[orange, thick,->] (2.58,-4.85) -- (2.58,-5.65);
    \end{tikzpicture}\\ 
     \begin{tikzpicture} 
        \node[anchor=north west,inner sep=0pt] at (0,0){\includegraphics[width=5.3cm]{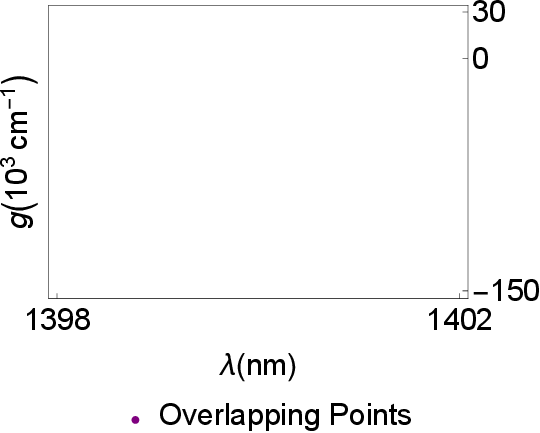}};
        \node[font=\sffamily\bfseries\normalsize] at (6ex,1.5ex) {(d)};
        \end{tikzpicture}
     \begin{tikzpicture} 
        \node[anchor=north west,inner sep=0pt] at (0,0){\includegraphics[width=5cm]{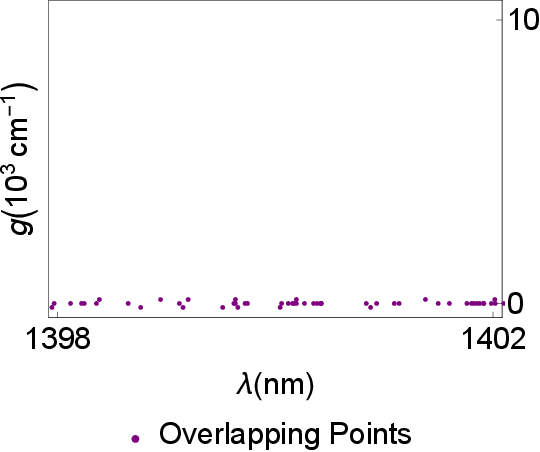}};
        \node[font=\sffamily\bfseries\normalsize] at (6ex,1.5ex) {(e)};
        \end{tikzpicture}\ 
     \begin{tikzpicture} 
        \node[anchor=north west,inner sep=0pt] at (0,0){\includegraphics[width=5cm]{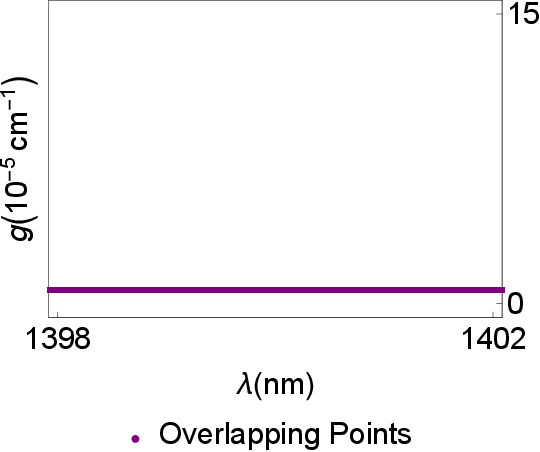}};
        \node[font=\sffamily\bfseries\normalsize] at (6ex,1.5ex) {(f)};
        \end{tikzpicture}
  \caption{(Color Online) In the graphs, spectral singularities are displayed over the $\lambda - g$ plane of the Plus Mode configuration corresponding to distinct $\mathtt{b}'$ values. The graphs in the upper panels correspond to the values $\mathtt{b}' = \infty~{\AA}$ (or $\mathtt{b} = 0~{\AA}^{-1}$), $\mathtt{b}' = 0.5~{\AA}$ and $\mathtt{b}' = 0.0005~{\AA}$ respectively, and each color corresponds to a different real zero component of the transfer matrix. Here, the red color belongs to the zeros of the $\mathbb{M}_{23}$, the blue color to the $\mathbb{M}_{33}$ and the green color to the $\mathbb{M}_{43}$ components. Graphs in the lower panel indicate the intersection points of the points shown in different colors in the upper panels. Notice that graphs are drawn using the data in (\ref{specification}). As can be seen from the lower graphs, only special values of $\mathtt{b}'$ (In our case $\mathtt{b}' = 0.5~{\AA}$ and $\mathtt{b}' = 0.0005~{\AA}$) allow laser beam exit from both sides of the slab. Also note the topological character of the arrangement of spectral singularity points. Please note that the colors presented here are unrelated to those shown in the previous table.} \label{figg3}
  \end{figure*}

\begin{figure*}[ht!]
    \centering
        \begin{tikzpicture} 
        \node[anchor=north west,inner sep=0pt] at (0,0){\includegraphics[width=5.4cm]{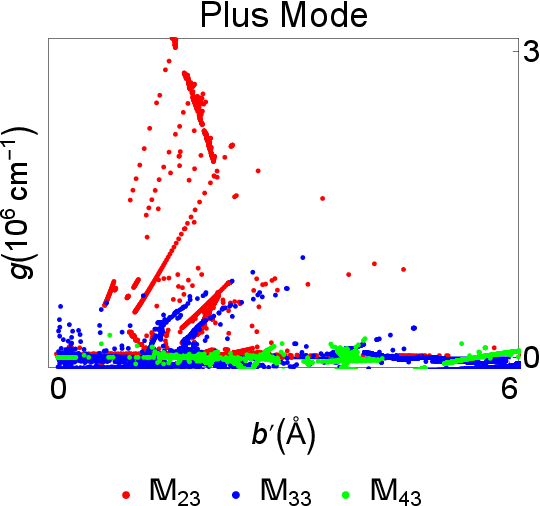}};
        \node[font=\sffamily\bfseries\normalsize] at (6ex, -1.5ex) {(a)}; \draw[orange, thick,->] (6.18,-4.0) -- (6.18,-4.80);
        \end{tikzpicture} ~
         \begin{tikzpicture}  
        \node[anchor=north west,inner sep=0pt] at (0,0){\includegraphics[width=5.4cm]{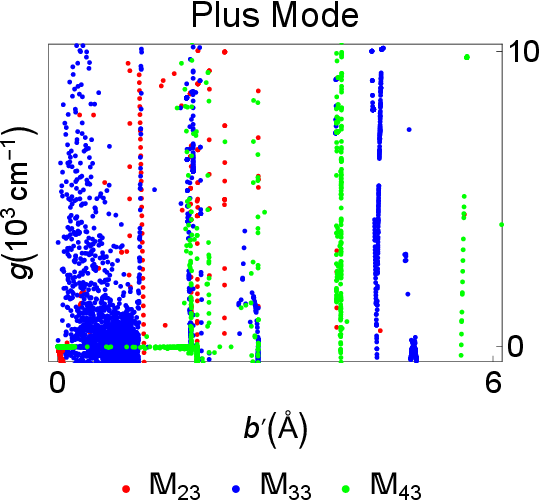}};
        \node[font=\sffamily\bfseries\normalsize] at (6ex, -1.5ex) {(b)}; 
    \end{tikzpicture}\\
      \begin{tikzpicture} 
        \node[anchor=north west,inner sep=0pt] at (0,0){\includegraphics[width=5.4cm]{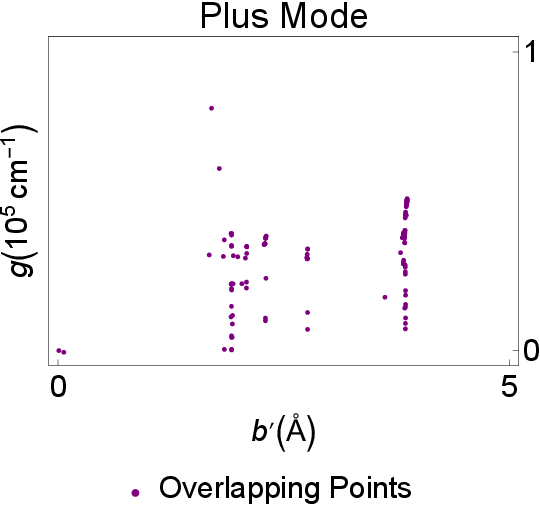}};
        \node[font=\sffamily\bfseries\normalsize] at (6ex, -1.5ex) {(c)}; 
        \end{tikzpicture} 
  \caption{(Color Online) In the graphs, spectral singularities are displayed over the $\mathtt{b}' - g$ plane of the Plus Mode configuration corresponding to resonance wavelength $\lambda = 1400~\textrm{nm}$. The graph in the upper panels corresponds to a different real zero components of the transfer matrix. Here, the red color belongs to the zeros of the $\mathbb{M}_{23}$, the blue color to the $\mathbb{M}_{33}$ and the green color to the $\mathbb{M}_{43}$ components. Graph in the lower panel indicate the intersection points of the points shown in different colors in the upper panels. As can be seen from the lower graph, only special values of $\mathtt{b}'$ allow laser beam exit from both sides of the slab, verifying the observation found in (\ref{figg3}). Also note the topological character of the arrangement of spectral singularity points.} \label{figg4}
\end{figure*}

\begin{figure*}[ht!]
    \centering
        \begin{tikzpicture} 
        \node[anchor=north west,inner sep=0pt] at (0,0){\includegraphics[width=5cm]{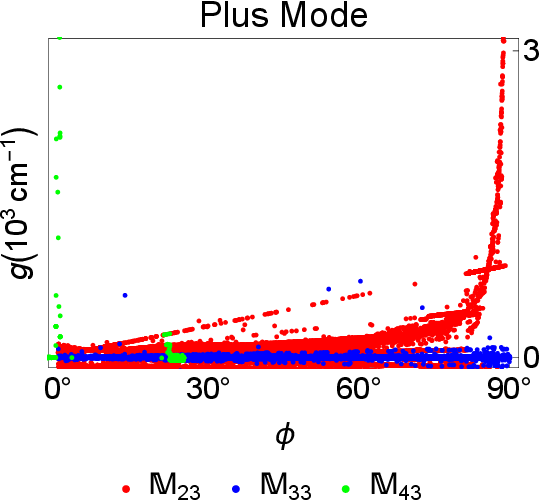}};
        \node[font=\sffamily\bfseries\normalsize] at (6ex, -1.5ex) {(a)}; \draw[orange, thick,->] (2.7,-5.0) -- (2.7,-5.6);
        \end{tikzpicture} ~~~
         \begin{tikzpicture}  
        \node[anchor=north west,inner sep=0pt] at (0,0){\includegraphics[width=5cm]{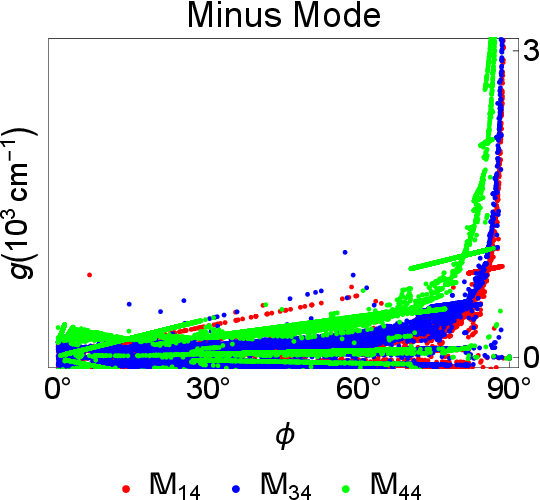}};
        \node[font=\sffamily\bfseries\normalsize] at (6ex, -1.5ex) {(b)}; \draw[orange, thick,->] (2.7,-5.0) -- (2.7,-5.6);
    \end{tikzpicture}\\
      \begin{tikzpicture} 
        \node[anchor=north west,inner sep=0pt] at (0,0){\includegraphics[width=5cm]{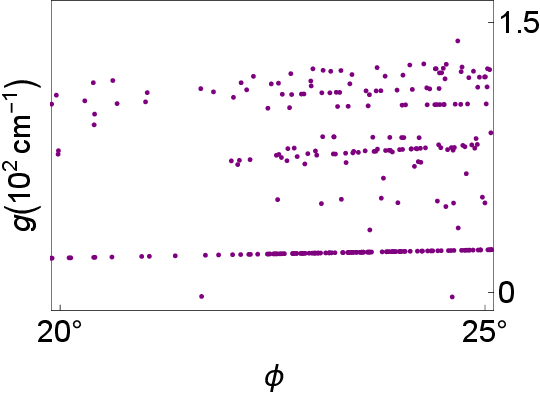}};
        \node[font=\sffamily\bfseries\normalsize] at (6ex, 1.5ex) {(c)}; 
        \end{tikzpicture} ~~~
              \begin{tikzpicture} 
        \node[anchor=north west,inner sep=0pt] at (0,0){\includegraphics[width=5cm]{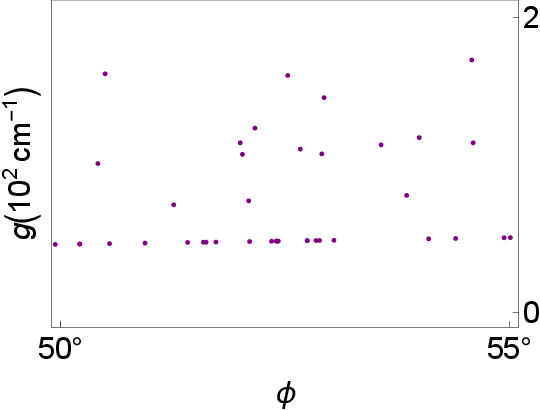}};
        \node[font=\sffamily\bfseries\normalsize] at (6ex, 1.5ex) {(d)}; 
        \end{tikzpicture} 
  \caption{(Color Online) In the figures, the locations of the spectral singularities belonging to the Plus and Minus Modes in the $\phi$-$g$ plane are shown. Here, the wavelength is $\lambda = 1400~\textrm{nm}$ and the $\mathtt{b}'$ value is $\mathtt{b}'= 0.05~{\AA}$. Each distinct color in the upper panels shows the real zeros of certain components of the transfer matrix. While the entire angle ranges are seen in the upper panels, the intersections of the spectral singularity points in the upper panels are seen in the lower ones. In other words, the spectral singularities belonging to the relevant laser configurations are the points in the lower panel. The topological character of the arrangement of the points is clearly seen.} \label{figg5}
\end{figure*}

$\mathcal{PT}$-symmetric TWS slab system is governed by several factors, including the gain coefficient $g$, the wavelength $\lambda$ of the incoming wave, distance between Weyl nodes $\mathtt{b}$, the incident angle $\phi$, the thickness of the slab $L$, and the material properties, represented by $\eta$. Typically, since the $\mathtt{b}$ value is a parameter defined within the Brillouin zone, we will utilize its spatial counterpart, denoted as $\mathtt{b}'$, i.e., $\mathtt{b} = 2 \pi / \mathtt{b}'$. The optimal conditions arise from the proper interdependence of these parameters. For simplicity, we choose TaAs as the TWS material, with its properties outlined in Eq.(\ref{specification}).  Since the lasing characteristic is primarily governed by the gain coefficient $g$, we will explore how other parameters influence the existence of spectral singularities by examining how the gain varies with changes in these parameters.The relationship between the gain coefficient $g$ and the wavelength $\lambda$ for Minus and Plus Mode types of lasers is illustrated in Figs. (\ref{figg1}) and (\ref{figg3}). The Figs.~ (\ref{figg2}) and (\ref{figg4}) examine the relationship between $g$ and the parameter $\mathtt{b}'$, and then, the effect of the incident wave angle $\phi$ on $g$ is discussed in the subsequent Fig.~(\ref{figg5}).

In Fig.~ (\ref{figg1}), the spectral singularity configurations for the Minus Mode are analyzed using three different values of $\mathtt{b}'$. The upper panel graphs show the real zeros of the $\mathbb{M}_{14}$, $\mathbb{M}_{34}$, and $\mathbb{M}_{44}$ components of the transfer matrix, represented in different colors. The lower panel graphs display the intersection points of these components from the upper panels. In summary, the lasing points for the Minus Mode are determined by the spectral singularity points shown at the bottom panels. As observed, achieving the laser threshold condition becomes more challenging at higher values of $\mathtt{b}'$. However, as the $\mathtt{b}'$-value decreases, the system exhibits more spectral singularity points, facilitating better lasing performance. Observe the horizontal alignment of the spectral singularity points in the graphs. This illustrates the topological nature of the spectral singularities, where the wavelength $\lambda$ may vary, but the gain coefficient remains robust against any changes.

Fig.~(\ref{figg2}) illustrates the variation of the gain coefficient $g$ as a function of $\mathtt{b}'$ for the Minus Mode laser type. The wavelength is set to the resonance wavelenth for TaAs $\lambda = 1400~\textrm{nm}$ and the incident angle is $\phi = 30^{\circ}$. On the left side of the figure, the real zeros of the $\mathbb{M}_{14}$, $\mathbb{M}_{34}$, and $\mathbb{M}_{44}$ components of the transfer matrix are displayed in different colors. On the right, the spectral singularity points of the Minus Mode, which occur due to the overlap of these components, are shown. As observed, multiple gain values at the same $g$ highlight that $\mathtt{b}'$ plays a role in enhancing topological robustness.

Fig.~(\ref{figg3}) displays the spectral singularity values corresponding to the real zero points of the relevant transfer matrix components $\mathbb{M}_{23}$, $\mathbb{M}_{33}$, and $\mathbb{M}_{43}$  for three different values of $\mathtt{b}'$ in the Plus Mode lasing configuration. Each component is represented by a different color. The upper panels show the graphs for $\mathtt{b} = 0~{\AA}^{-1}$, $\mathtt{b}'= 0.5~{\AA}$, and $\mathtt{b}'= 0.0005~{\AA}$, respectively. The lower panels present the intersection points of the different components from the upper panels. These intersection points correspond to the Plus Mode lasing threshold values, which are defined by the spectral singularity points shown below. As observed, for $\mathtt{b}= 0~{\AA}^{-1}$, the relevant components of the transfer matrix do not intersect, preventing the system from lasing. However, when $\mathtt{b}$ is reduced to a non-zero value, the lasing points increase and interestingly, they exhibit behavior that maintains certain gain values. This reinforces the idea that the system is topologically robust, as it can lase at the same gain coefficient across multiple wavelength values.

Fig.~(\ref{figg4}) illustrates the variation of the gain coefficient $g$ with respect to $\mathtt{b}'$ in the Plus Mode, with the graphs using $\lambda = 1400~\textrm{nm}$ and $\phi = 30^{\circ}$ as parameters. In the upper panel(s), the real zeros of the relevant transfer matrix components for two different ranges are shown. The intersections of these points are displayed in the lower panel. As seen, the intersection points at the bottom correspond to the spectral singularity points we are seeking, which define the Plus Mode lasing configuration. Notably, for the same $\mathtt{b}'$-values, multiple gain values are observed. This highlights the topological structure of the system as it depends on $\mathtt{b}'$.

Finally, we examine how the gain value is influenced by the incident angle $\phi$ in both the Plus and Minus Modes. To do this, we plot the variation of $g$ with $\phi$, using the same values for $\lambda = 1400~\textrm{nm}$ and $\mathtt{b}'= 0.05~{\AA}$ as in Fig.~(\ref{figg5}). In the upper panels, the real zeros of the relevant components of the transfer matrix are shown in three different colors. The intersections of these components are displayed in the lower panels. Surprisingly, lasing is not observed at every angle for the Plus Mode. However, for the Minus Mode, the system exhibits lasing at many different angle values. In the panels below, the behavior of the intersections within specific angle ranges is shown. As observed, lasing configurations corresponding to the same gain value can be achieved at various angle values. This reinforces the idea that $g$ exhibits a topological character that is dependent on $\phi$ in $\mathcal{PT}$-symmetric TWS systems.

\section{Effect of  $\mathcal{PT}$ symmetry on TWS Laser Types}
\label{S8}

In this section, we will analyze the influence of  $\mathcal{PT}$ symmetry on the location of spectral singularities, as well as its effect on the different types of topological lasers and the conditions required for their formation. To achieve this, we modify the loss component in the TWS system to a gain, meaning $\fn_{loss}= \fn^{*} \rightarrow \fn_{gain}=\fn$. Thus, our $\mathcal{PT}$-symmetric TWS system will be converted into a TWS system with a thickness of $2L$, containing only gain. To understand the influence of $\mathcal{PT}$ symmetry, we can compare the relevant parameters of the lasing configuration with those of the $\mathcal{PT}$-symmetric TWS system. The key indicator in this comparison is, in fact, the values of $g$.

\begin{figure*}[ht!]
    \centering
        \begin{tikzpicture} 
        \node[anchor=north west,inner sep=0pt] at (0,0){\includegraphics[width=5cm]{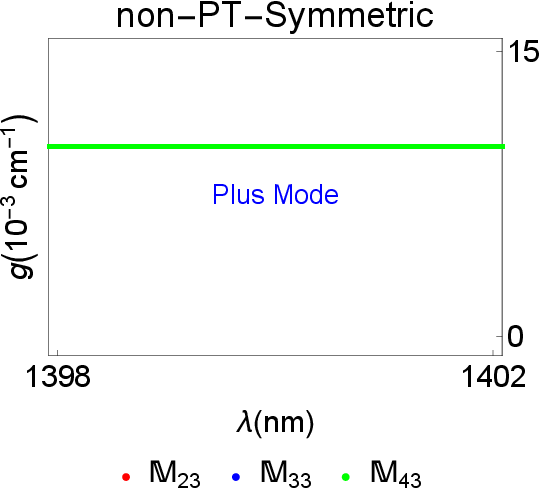}};
        \node[font=\sffamily\bfseries\normalsize] at (6ex, -1.3ex) {(a)}; \draw[orange, thick,->] (5.48,-1.9) -- (6.48,-1.9);
        \end{tikzpicture} ~~~
         \begin{tikzpicture}  
        \node[anchor=north west,inner sep=0pt] at (0,0){\includegraphics[width=5cm]{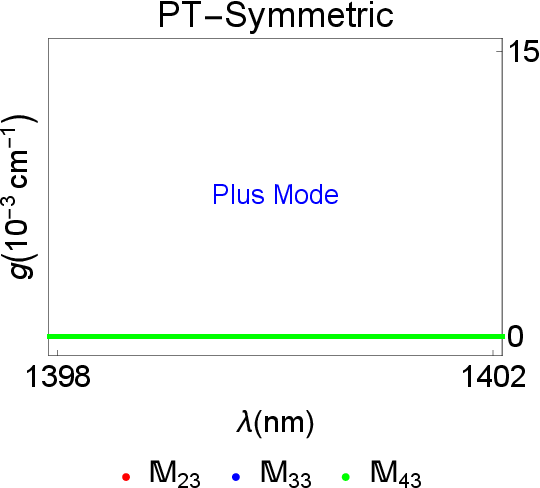}};
        \node[font=\sffamily\bfseries\normalsize] at (6ex, -1.3ex) {(b)}; 
    \end{tikzpicture}\\
            \begin{tikzpicture} 
        \node[anchor=north west,inner sep=0pt] at (0,0){\includegraphics[width=5cm]{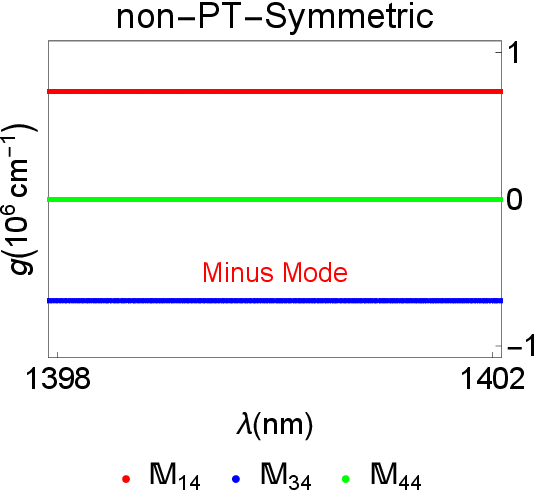}};
        \node[font=\sffamily\bfseries\normalsize] at (6ex, -1.3ex) {(c)}; \draw[orange, thick,->] (5.48,-1.9) -- (6.48,-1.9);
        \end{tikzpicture} ~~~
         \begin{tikzpicture}  
        \node[anchor=north west,inner sep=0pt] at (0,0){\includegraphics[width=5cm]{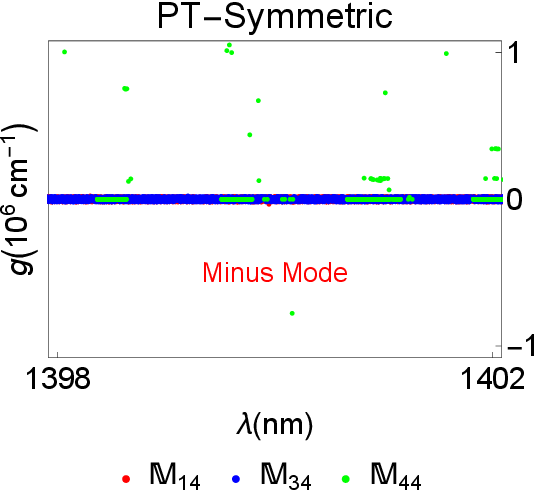}};
        \node[font=\sffamily\bfseries\normalsize] at (6ex, -1.3ex) {(d)}; 
    \end{tikzpicture}
  \caption{(Color Online) The figure illustrates the impact of $\mathcal{PT}$ symmetry under both Plus and Minus Mode configurations, with a $\mathtt{b}'$ value of $0.0005~{\AA}$. As shown, imposing $\mathcal{PT}$ symmetry leads to a significant reduction in the lasing gain. When $\mathcal{PT}$ symmetry is removed, the system continues to lase at higher gain values in Plus Mode, but no lasing occurs in Minus Mode.} \label{figg6}
\end{figure*}

Figure~(\ref{figg6}) illustrates the impact of $\mathcal{PT}$ symmetry on the laser performance of the TWS system. It demonstrates that introducing $\mathcal{PT}$ symmetry to a TWS system significantly reduces the laser threshold gain. This outcome aligns with known results in conventional systems and, as shown, holds true in a topological context as well. Specifically, the application of $\mathcal{PT}$ symmetry leads to a reduction in the required gain by roughly a factor of 1000. This reduction not only enhances the laser performance but also facilitates more precise control over the gain and improves overall system efficiency. The figure also highlights that, in the absence of $\mathcal{PT}$ symmetry and with the corresponding parameters, the system does not exhibit lasing behavior in the Minus Mode. However, when $\mathcal{PT}$ symmetry is applied, the system is able to lase under the same set of parameter values. These observations emphasize the critical role of incorporating $\mathcal{PT}$-symmetric structures in laser systems for achieving enhanced performance.

\section{Axion-Induced Current $\vec{\mathcal{J}}_{\theta}$ and its Topological Character}
\label{S81}

Another key aspect to address in this context is the axion-induced current. In the expression for current density in Maxwell's third equation, there is not only the free current but also an additional induced current arising from the axion term $\theta$, which imparts the topological character to the system. We refer to this current as the axion-induced current and denote it by $\vec{\mathcal{J}}_{\theta}$. The origin of this current lies in the semimetallic nature of the TWS medium, and it manifests not only within the bulk of the material but also on its surface. Therefore, the existence and detection of this axion-induced current are crucial for understanding the system's topological properties.  If we examine the expression for the current density $\vec{\mathcal{J}}_{\theta}$, it appears as a vectorial relation as follows:
\be
\vec{\mathcal{J}}_{\theta} = -\beta\,\vec{\mathtt{b}}\times\vec{\cE}.
\ee
This current oscillates harmonically over time. The time-independent expression for this current in gain or loss medium can be clearly written as follows,
\be
\vec{\mathcal{J}}_{\theta}^{\mathfrak{u}} = -\frac{\mathtt{b} \beta e^{i\fK \mathbf{x}\tan\phi}}{2}  \left\{ i(\mathcal{F}_+^{\mathfrak{u}} - \mathcal{G}_+^{\mathfrak{u}})\hat{e}_x + (\mathcal{F}_+^{\mathfrak{u}} + \mathcal{G}_+^{\mathfrak{u}})\hat{e}_y\right\}.
\ee
Here, the symbol $\mathfrak{u}$ denotes the material's nature, representing either gain or loss, i.e., $\mathfrak{u} \in [\textrm{gain}, \textrm{loss}]$. In other words, the current flowing through the material will exhibit distinct behaviors depending on whether the material is in a gain or loss regime. It is evident that this current flows within a plane perpendicular to the direction of the material's structural distribution, which is similar to the Hall current. However, the magnitude of this current is dependent on the specific conditions of each spectral singularity case corresponding to the associated laser Mode. When the relevant conditions for the Plus Mode, i.e., $A_{1}^{(+)} = A_{1}^{(-)} = C_{1}^{(-)} = C_{3}^{(+)} = A_{3}^{(-)} = C_{3}^{(-)} = 0$, are applied such that $A_{3}^{(+)} = \mathbb{M}_{13} \,C_1^{+} $ , it simplifies to the following form,
\small
\begin{flalign}
\vec{\mathcal{J}}_{\theta}= \frac{i \mathtt{b} \beta  C_1^{+} e^{i\fK \mathbf{x}\tan\phi} }{2} \left\{\begin{array}{cc}
   \left[ i \cos(\fK_+^{g} \mathbf{z}) + \mathfrak{a}_-^{g} \sin(\fK_+^{g} \mathbf{z}) \right] (\hat{e}_y + i \hat{e}_x) + \sin (\fK_-^{g} \mathbf{z}) \mathfrak{b}^{g}  (\hat{e}_y - i \hat{e}_x)& {\rm for}~~ 0 < \mathbf{z} < 1,\\
  \left[  i \cos[\fK_+^{\ell} (\mathbf{z}-2)] - \mathfrak{a}_+^{\ell} \sin[\fK_+^{\ell} (\mathbf{z}-2)] \right] \mathbb{M}_{13} e^{2i\fK}(\hat{e}_y + i \hat{e}_x) + \sin [\fK_- ^{\ell}(\mathbf{z}-2)] \mathbb{M}_{13}  \mathfrak{b}^{\ell}  e^{2i\fK} (\hat{e}_y - i \hat{e}_x)& {\rm for}~~ 1 < \mathbf{z} < 2 ,
    \end{array}\right. \notag
\end{flalign}
\normalsize
where the following identifications are employed for convenience
\begin{align}
\mathfrak{a}_{\pm}^{g/\ell} := \frac{\mu(1 \pm \sigma_{+})}{ \tilde{\fn}_{+}^{g/\ell}}, \qquad  \mathfrak{b}^{g/\ell} := \frac{\mu \sigma_{+}}{ \tilde{\fn}_{-}^{g/\ell}}, \qquad \fK_{\pm}^{g/\ell} :=  \fK \tilde{\fn}_{\pm}^{g/\ell} \notag
\end{align}
and the superscript $g/\ell$ implies the gain and loss parts of TWS medium with $\tilde{\fn}_{\pm}^{g} := \tilde{\fn}_{\pm}$ and $\tilde{\fn}_{\pm}^{\ell} := \tilde{\fn}_{\pm}^{*}$. Notice that the gain part is determined by the range $\mathbf{z} \in [0, 1]$, whereas the loss part is governed by $\mathbf{z} \in [1, 2]$. Likewise, under the conditions for the Minus Mode configuration, i.e. $A_{1}^{(+)} = A_{1}^{(-)} = C_{1}^{(+)} = C_{3}^{(+)} = A_{3}^{(+)} = C_{3}^{(-)} = 0$ such that $A_{3}^{(-)} = \mathbb{M}_{24} \,C_1^{-} $ , the induced current can be expressed as follows:
\small
\begin{flalign}
\vec{\mathcal{J}}_{\theta}= \frac{i \mathtt{b} \beta  C_1^{-} e^{i\fK \mathbf{x}\tan\phi} }{2}  \left\{\begin{array}{cc}
   \left[ i \cos(\fK_-^{g} \mathbf{z}) + \mathfrak{c}_-^{g} \sin(\fK_-^{g} \mathbf{z}) \right] (\hat{e}_y - i \hat{e}_x) + \sin (\fK_+^{g} \mathbf{z}) \mathfrak{d}^{g}  (\hat{e}_y + i \hat{e}_x)& {\rm for}~~ 0 < \mathbf{z} < 1,\\
  \left[  i \cos[\fK_-^{\ell} (\mathbf{z}-2)] - \mathfrak{c}_+^{\ell} \sin[\fK_-^{\ell} (\mathbf{z}-2)] \right] \mathbb{M}_{24} e^{2i\fK}(\hat{e}_y - i \hat{e}_x) + \sin [\fK_+ ^{\ell}(\mathbf{z}-2)] \mathbb{M}_{24}  \mathfrak{d}^{\ell}  e^{2i\fK} (\hat{e}_y + i \hat{e}_x)& {\rm for}~~ 1 < \mathbf{z} < 2 ,
    \end{array}\right. \notag
\end{flalign}
\normalsize
where we identify the following quantities
\begin{align}
\mathfrak{c}_{\pm}^{g/\ell} := \frac{\mu(1 \pm \sigma_{-})}{ \tilde{\fn}_{-}^{g/\ell}}, \qquad  \mathfrak{d}^{g/\ell} := \frac{\mu \sigma_{-}}{ \tilde{\fn}_{+}^{g/\ell}}. \notag
\end{align}

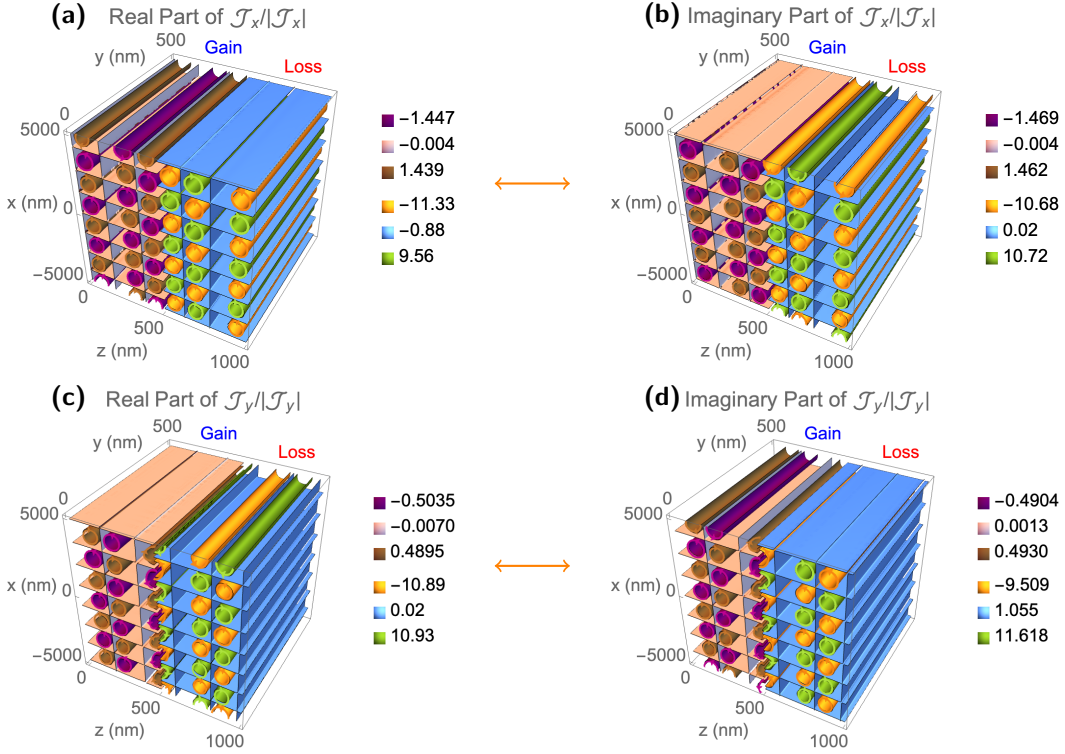
\begin{figure}[ht!]
    \centering
        \begin{tikzpicture} 
        \node[anchor=north west,inner sep=0pt] at (0,0){\includegraphics[width=6cm]{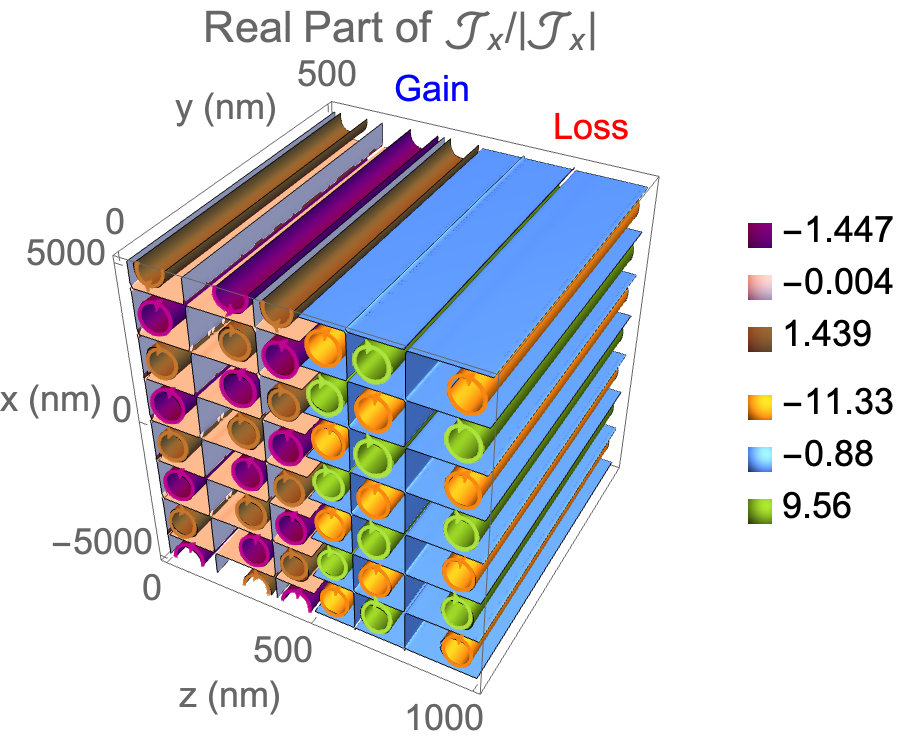}};
        \node[font=\sffamily\bfseries\normalsize] at (6ex, -1.3ex) {(a)}; \draw[orange, thick,<->] (6.48,-2.4) -- (7.48,-2.4);
        \end{tikzpicture} ~~~
         \begin{tikzpicture}  
        \node[anchor=north west,inner sep=0pt] at (0,0){\includegraphics[width=6cm]{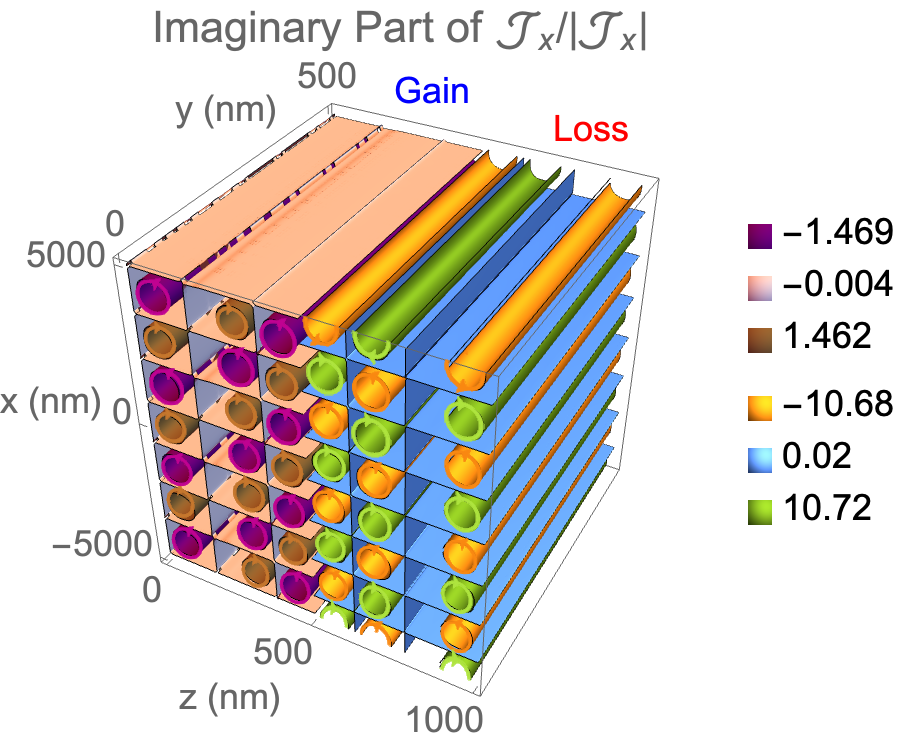}};
        \node[font=\sffamily\bfseries\normalsize] at (5ex, -1.3ex) {(b)}; 
    \end{tikzpicture}\\
            \begin{tikzpicture} 
        \node[anchor=north west,inner sep=0pt] at (0,0){\includegraphics[width=6cm]{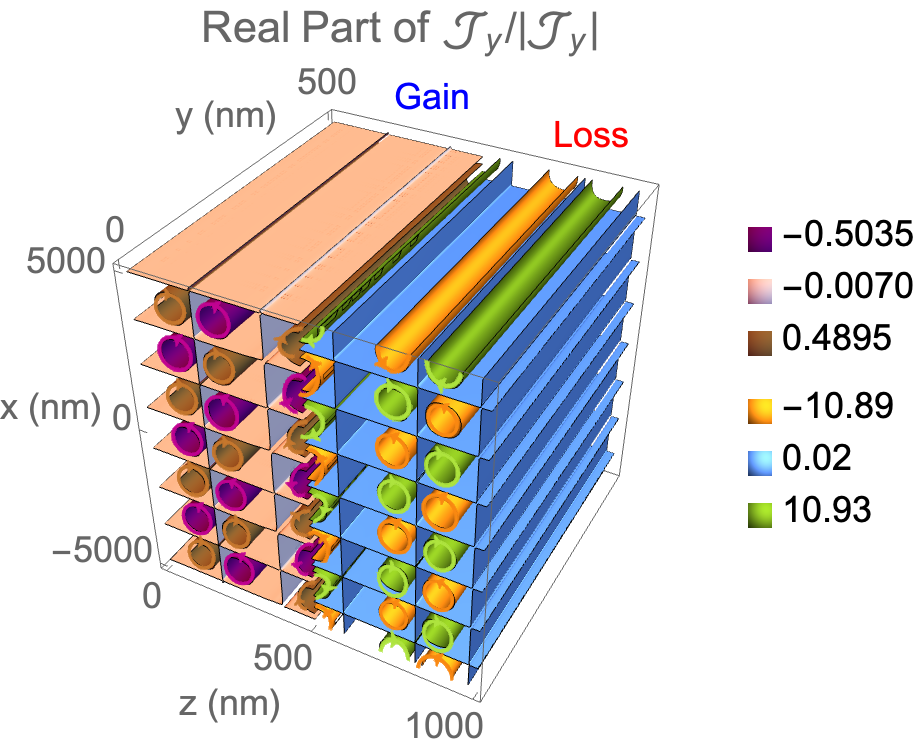}};
        \node[font=\sffamily\bfseries\normalsize] at (6ex, -1.3ex) {(c)}; \draw[orange, thick,<->] (6.48,-2.4) -- (7.48,-2.4);
        \end{tikzpicture} ~~~
         \begin{tikzpicture}  
        \node[anchor=north west,inner sep=0pt] at (0,0){\includegraphics[width=6cm]{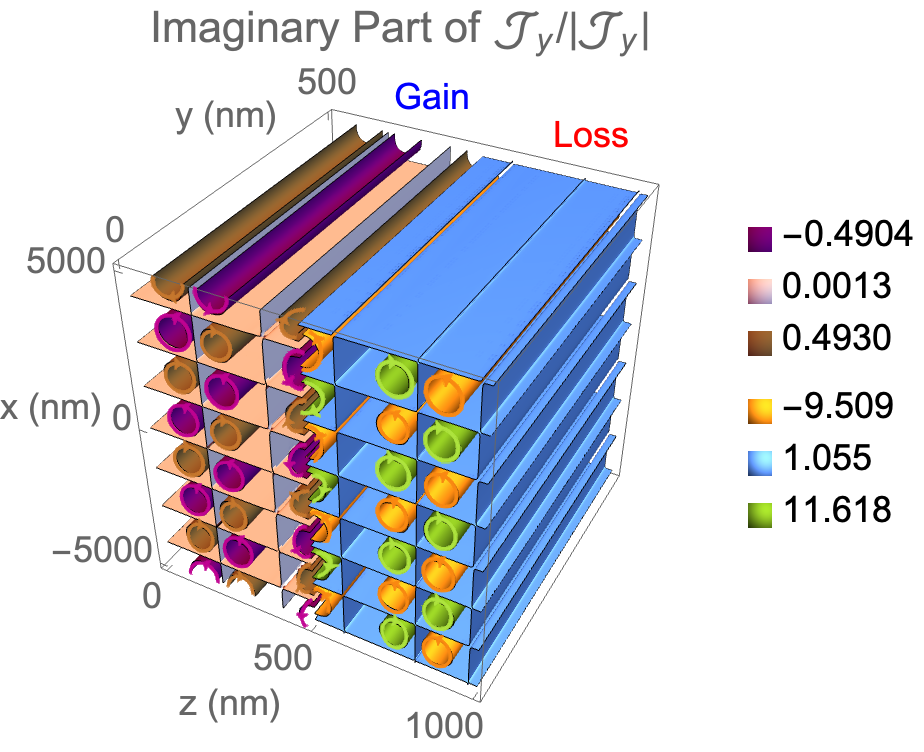}};
        \node[font=\sffamily\bfseries\normalsize] at (5ex, -1.3ex) {(d)}; 
    \end{tikzpicture}
  \caption{(Color Online) Axion-induced currents observed in the plus mode spectral singularity configuration of a $\mathcal{PT}$-symmetric TWS slab. The distinct behavior of the currents in the gain and loss regions of the material is evident. The upper row shows the real and imaginary parts of the $x$-component of the current, while the lower row displays the behavior of the $y$-component. These currents, which are axion-induced Hall currents, exhibit a circular pattern.} \label{figg7}
\end{figure}

\begin{figure}[ht!]
    \centering
        \begin{tikzpicture} 
        \node[anchor=north west,inner sep=0pt] at (0,0){\includegraphics[width=6cm]{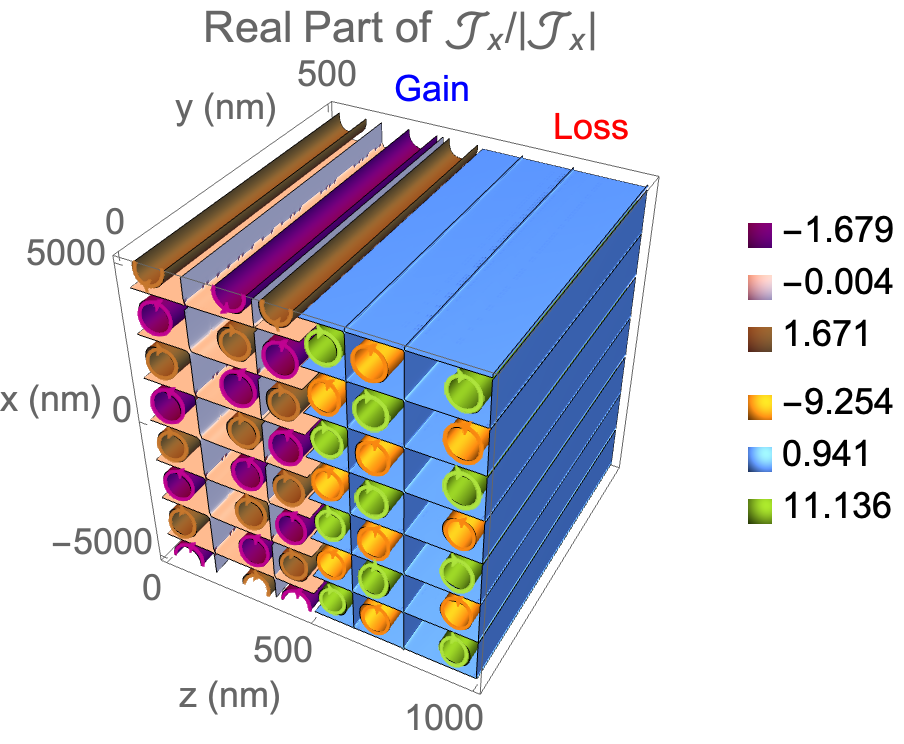}};
        \node[font=\sffamily\bfseries\normalsize] at (6ex, -1.3ex) {(a)}; \draw[orange, thick,<->] (6.48,-2.4) -- (7.48,-2.4);
        \end{tikzpicture} ~~~
         \begin{tikzpicture}  
        \node[anchor=north west,inner sep=0pt] at (0,0){\includegraphics[width=6cm]{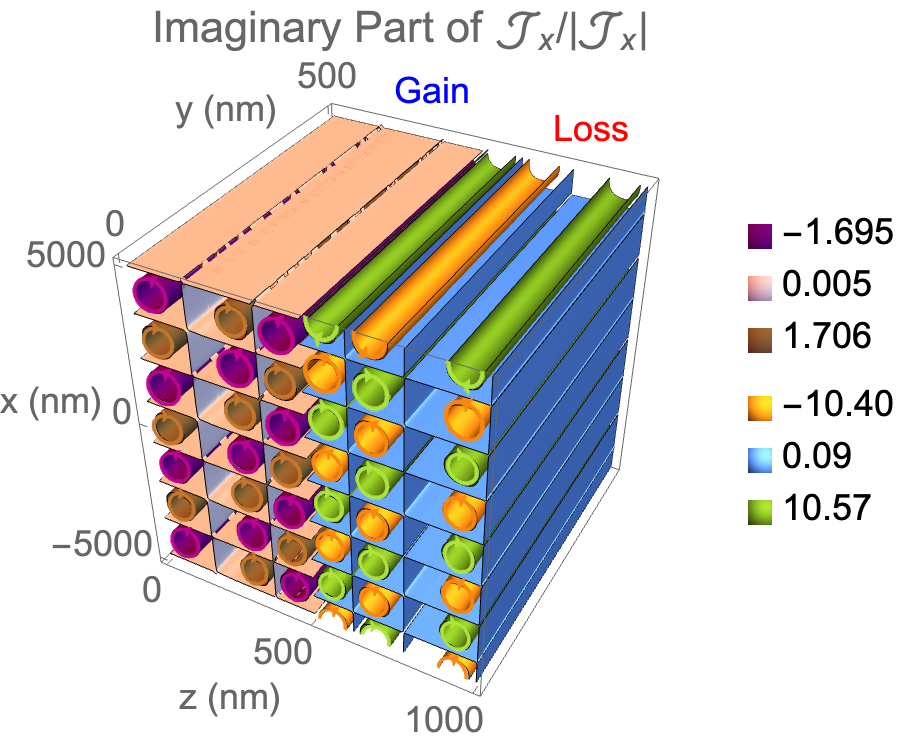}};
        \node[font=\sffamily\bfseries\normalsize] at (5ex, -1.3ex) {(b)}; 
    \end{tikzpicture}\\
            \begin{tikzpicture} 
        \node[anchor=north west,inner sep=0pt] at (0,0){\includegraphics[width=6cm]{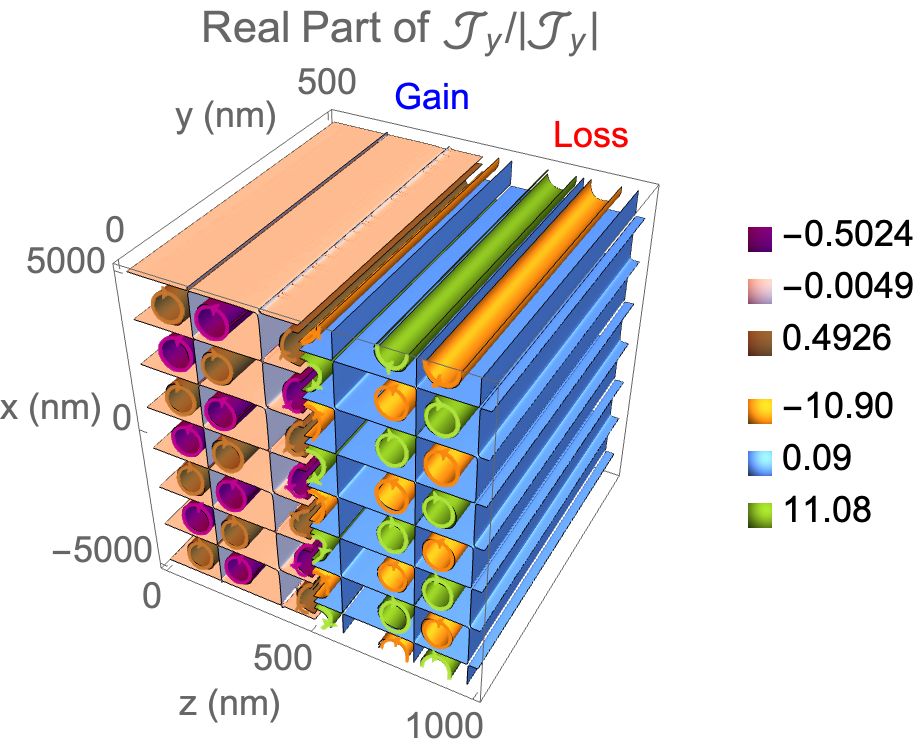}};
        \node[font=\sffamily\bfseries\normalsize] at (6ex, -1.3ex) {(c)}; \draw[orange, thick,<->] (6.48,-2.4) -- (7.48,-2.4);
        \end{tikzpicture} ~~~
         \begin{tikzpicture}  
        \node[anchor=north west,inner sep=0pt] at (0,0){\includegraphics[width=6cm]{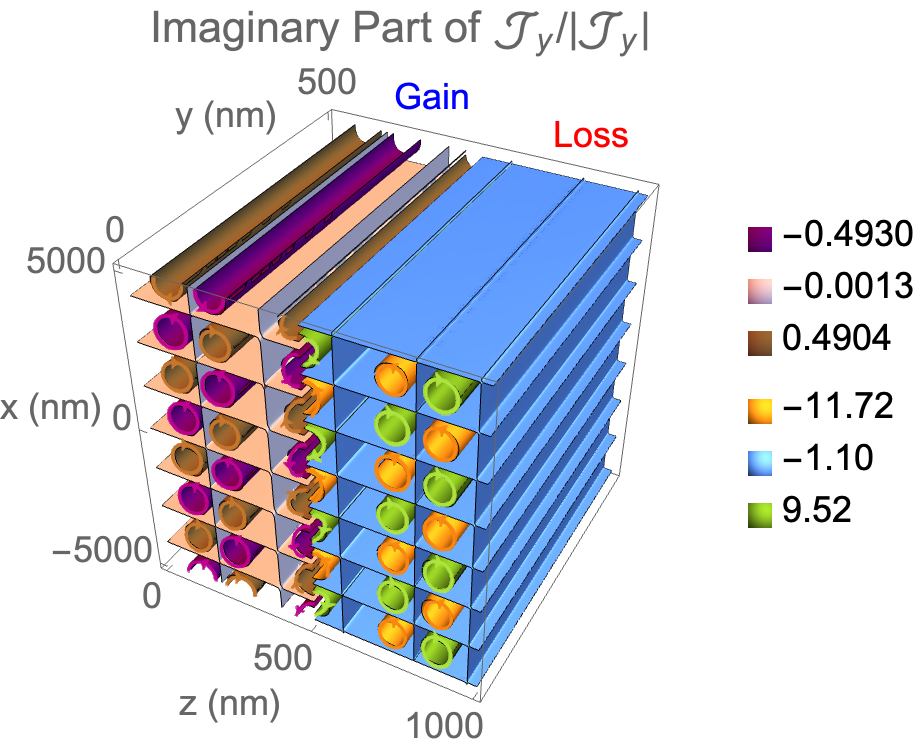}};
        \node[font=\sffamily\bfseries\normalsize] at (5ex, -1.3ex) {(d)}; 
    \end{tikzpicture}
  \caption{(Color Online) Axion-induced currents observed at the minus-mode spectral singularity in a $\mathcal{PT}$-symmetric TWS slab. The currents' behavior in the gain and loss regions of the material is clearly distinguished. The upper row presents the real and imaginary components of the current along the $x$-axis, while the lower row shows the behavior of the current along the $y$-axis. These axion-driven Hall currents exhibit a circular shape.} \label{figg8}
\end{figure}

Figures~(\ref{figg7}) and (\ref{figg8}) present the real and imaginary parts of induced current distributions for the plus and minus mode configurations, respectively, within and on the surfaces of the  $\mathcal{PT}$-symmetric TWS medium, at the moments when spectral singularities arise. These current patterns are similar to those reported in Ref.~\cite{sevval}, which supports the idea that TWS environments facilitate the formation of cyclotron-type currents due to the axion terms they contain. The spectral singularity parameters used in these plots are specified as $\mathtt{b}'= 3~{\AA}$, $\lambda = 1405.265$, and $g = 5.763\times 10^{-5}~\textrm{cm}^{-1}$. Notably, higher values of $\mathtt{b}'$ are chosen to obtain more distinct current distributions. As the value of $\mathtt{b}'$ is reduced, the number of circular patterns observed in the graphs decreases. Similarly, increasing the gain value shifts the positions of these circular patterns disorderly and anomalously along the $x$-axis, which is why the smallest possible gain values were utilized in this study.

From the graphs, it is evident that the axion-induced currents exhibit a fully cyclotron-like behavior, flowing in specific directions. This behavior is also an indication that the topological character of the system is preserved while the induced currents circulate in the specified directions. In other words, the existence of these currents is actually a demonstration of the topological behavior originating from the $\mathtt{b}'$-term in the TWS environment. The topological symmetry here manifests itself in cyclotron-type shapes. Positive currents rotate in the clockwise direction, while negative currents rotate counterclockwise. The flow directions of these currents demonstrate a well-defined symmetry: in the gain and loss regions, the rotations are perfectly synchronized, complementing each other. Interestingly, in the loss medium, the rotation directions of currents in the plus mode are diametrically opposite to those in the minus mode, indicating that the plus and minus mode configurations produce contrasting current flow patterns in the lossy medium.

Lastly, these induced currents are confined to the $xy$-plane, while the electromagnetic wave propagates along the $z$-axis, thereby manifesting as a Hall current.

\section{Concluding Remarks}
\label{S9}

Our study has yielded significant insights into the interaction between $\mathcal{PT}$-symmetric TWS slab and electromagnetic waves, with a particular focus on transforming this interaction into a topological laser through the controlled manipulation of system parameters. By analyzing the conditions for the existence of spectral singularities, we identified the necessary criteria for constructing an effective topological laser. A key feature of the TWS medium is its axion term, which induces a polarization rotation of the wave due to magneto-optic effects upon interaction with the electromagnetic wave. This leads to Kerr and Faraday rotations, which occur when the wave is incident in the TE mode. These findings contribute to the development of new strategies for designing topologically protected laser systems.

Our analysis explored the impact of $\mathcal{PT}$ symmetry and its potential advantages in the formation of topological lasers. We found that the required gain for the topological laser based on $\mathcal{PT}$ symmetry can be significantly reduced, a result that aligns with predictions from previous studies. However, the more surprising discovery was that the TWS material medium, through Kerr and Faraday rotations, effectively increases the system's dimension, thereby enriching the diversity of topological lasers. For the first time, we demonstrated that a TWS medium can give rise to twelve distinct types of topological lasers. In particular, we examined the unique characteristics of these lasers, with a special focus on the ``Plus" and ``Minus" mode configurations.

Another important finding of our study is the connection between the scattering properties of the $\mathcal{PT}$-symmetric TWS medium and the Schrödinger equation. This relationship provides a valuable framework for investigating the non-Hermitian effects within such systems. As we initially discussed, the pathway to generating a topological laser in these environments involves analyzing the spectral singularity points within the non-Hermitian system. These points correspond to real energy values in the complex potential of the $\mathcal{PT}$-symmetric TWS medium, which in turn relate to the zeros of certain components of the transfer matrix at real $k$-values. In our study, we have clearly identified these points for both Plus and Minus mode configurations. Furthermore, we have demonstrated the critical role of the $\mathtt{b}$-term in the formation of spectral singularities, which are essential for the development of topological symmetries in these systems.

Our study also uncovered several novel findings, including the observation that the currents induced by the axion term in the $\mathcal{PT}$-symmetric TWS medium exhibit topologically protected and circular shapes, as revealed by the analysis of spectral singularity points. We examined the behavior of these currents in both the gain and loss regions, highlighting how the currents rotate during flow. Notably, the rotational states of the currents are influenced by the laser output states, with the Plus and Minus modes manifesting in completely opposite configurations in the loss section. This demonstrates that such materials inherently generate distinctive current patterns while facilitating the creation of a topological laser under $\mathcal{PT}$ symmetry.

As a final remark, topological devices are commonly characterized by phase diagrams that depict topological phase transitions within the parameter space. Although such diagrams are not included in the present study, we acknowledge their significance and intend to incorporate them in future work to provide a clearer understanding of how topological laser behavior depends on system parameters.

In summary, by carefully tuning the system parameters based on the insights gained from our study, it will be possible to develop a highly efficient topological laser. One of the key contributions of our work is the understanding of how the $\mathtt{b}$-term influences the topological symmetry of the system, providing a foundation for future research in this area. The significance and role of the $\mathtt{b}$-term remain a focal point of current research, and our findings contribute to the ongoing active discussions in the literature.\\[6pt]


\appendix
\section{Appendix}\label{boundarycond}

\subsection{A. Spectral Singularities: A Brief Overview}\label{spsing}

\textbf{Definition and Concept:}
A spectral singularity is a mathematical and physical concept that arises in the spectral analysis of non-Hermitian operators, particularly in complex scattering potentials. In essence, it corresponds to a real-energy pole of the resolvent operator (Green's function), and physically, it marks a critical point where the scattering amplitude becomes divergent while the potential remains reflectionless.

First introduced by M. A. Naimark in the 1950s in the context of the spectral theory of non-self-adjoint differential operators, the concept gained renewed attention through the work of Ali Mostafazadeh in the early 2000s, who connected it with optical systems and PT-symmetric quantum mechanics.

\textbf{Mathematical Formulation:} 
Let $H = -\frac{d^2}{dx^2} + V(x)$ be a one-dimensional Schrödinger operator with a complex-valued potential $V(x)$, defined on the real line. A spectral singularity occurs at a real energy $E = k^2 > 0$ for which:
\begin{itemize}
  \item The Jost solutions (asymptotically behaving like $e^{\pm ikx}$) become linearly dependent.
  \item The transmission and reflection coefficients become undefined due to divergence.
  \item The scattering matrix becomes non-diagonalizable at that energy.
\end{itemize}

This behavior implies that the system supports purely outgoing waves at the spectral singularity energy, analogous to zero-width resonances or threshold lasing conditions.

\textbf{Physical Interpretation:}
In optics and photonics, spectral singularities are associated with coherent perfect absorption (CPA) and lasing phenomena. For instance:
\begin{itemize}
  \item When a complex refractive index profile (e.g., in PT-symmetric media) supports a spectral singularity, the system may lase at the corresponding energy (or frequency) without an external cavity — this is termed a threshold laser mode.
  \item Conversely, the time-reversed counterpart of a spectral singularity corresponds to a CPA mode, where incident coherent waves are perfectly absorbed.
\end{itemize}

These effects have been exploited in designing non-Hermitian optical devices, unidirectional invisibility structures, and hybrid laser-absorber systems.

\textbf{Relevance in Non-Hermitian Physics:}
Spectral singularities are hallmark features of non-Hermitian systems and play a key role in the study of:
\begin{itemize}
  \item PT-symmetric quantum mechanics, where gain and loss are balanced.
  \item Topological photonics, where singularities signal topological transitions.
  \item Resonance scattering theory, especially for complex potentials.
\end{itemize}

They also challenge conventional assumptions in quantum mechanics by showing that even in absence of bound states or standard resonances, novel spectral features can arise purely due to non-Hermiticity.

\textbf{Summary:}
In summary, spectral singularities represent non-Hermitian analogues of resonance poles located on the real energy axis. They are closely tied to physical phenomena such as threshold lasing and perfect absorption, and their presence signals a critical change in the spectral and scattering behavior of the system. As such, they provide both theoretical depth and practical functionality in the design of novel quantum and photonic systems.

\subsection{B. Modified Maxwell Equations}\label{modmax}

In the low energy limit of a TWS, spatially varying axion term plays a significant role in determining its electromagnetic response. The full action of the corresponding TWS slab system is described by the sum of conventional and axionic terms as $S = S_0 + S_{\theta}$ as follows
  \begin{align}
  S_0 &= \int \left\{-\frac{1}{4\mu_0}F_{\mu\nu}F^{\mu\nu} +\frac{1}{2}F_{\mu\nu} \mathcal{P}^{\mu\nu} - J^{\mu}A_{\mu}\right\} d^3x\,dt,\notag\\
  S_{\theta} &= \frac{\alpha}{8\pi\mu_0}\int \left\{\theta(\vec{r}, t)\varepsilon^{\mu\nu\alpha\beta}F_{\mu\nu}F_{\alpha\beta}\right\} d^3x\,dt,\notag
  \end{align}
where $\mathcal{P}^{\mu\nu}$ tensor represents the electric polarization and magnetization respectively by $\mathcal{P}^{0i} = cP^{i}$ and $\mathcal{P}^{ij} =-\varepsilon^{ijk} M_{k}$. Space and time dependent axion term is given by $\theta(\vec{r}, t) = 2 \vec{\mathtt{b}}\cdot\vec{r}-2\mathtt{b}_0t$, where $\vec{\mathtt{b}}$ and $\mathtt{b}_0$ denote the separation of nodes in momentum and energy space respectively. In our case, $\mathtt{b}_0$ is set to be zero, since Weyl nodes are assumed to share the same chemical potential. If the action is varied with respect to $A_\mu$, the following equations of motion is obtained
  \be
  -\frac{1}{\mu_0}\partial_{\nu}F^{\mu\nu} + \partial_{\nu}\mathcal{P}^{\mu\nu} + \frac{\alpha}{2\pi\mu_0}\varepsilon^{\mu\nu\alpha\beta}\partial_{\nu}\left(\theta F_{\alpha\beta}\right) = J^{\mu}
  \ee
It is obvious that expanding this equation yields the modified Maxwell equations given by Eqs.~\ref{equation1} and \ref{equation2} in the presence of axion field term.

\subsection{C. Computation of Conductivities}\label{conduct}

To calculate the longitudinal $\sigma_{yy}$ and transverse $\sigma_{yx}$ conductivities of a TWS, we adopt the approaches given in \cite{axion3, conductiv1}. We consider the simplest case with only two nodes located at $+\vec{\mathtt{b}}$ and $-\vec{\mathtt{b}}$, where $\vec{\mathtt{b}} = \mathtt{b}\,\widehat{e}_{z}$ in our case. Near the Weyl nodes, the linearized Hamiltonian is given by
\be
H(\vec{k}) = \pm \hbar v_{F} \vec{\sigma}\cdot (\vec{k}\pm\vec{\mathtt{b}}).\notag
\ee
where $v_{F}$ is the Fermi velocity, and $\vec{\sigma} = (\sigma_x, \sigma_y, \sigma_z)$ is the vector whose components are the Pauli matrices. Therefore, conductivity $\sigma_{\alpha\beta}$ is obtained from Kubo formula as follows
\be
\sigma_{\alpha\beta} (\omega) = \frac{i}{\omega}\lim_{q\rightarrow 0}\Pi_{\alpha\beta}(q, \omega)\notag
\ee
In the absence of diamagnetic term, the polarization function $\Pi_{\alpha\beta}(q, \omega)$ is given by the current-current correlation function
\be
\Pi_{\alpha\beta}(q, i\omega_n) =  \frac{-1}{\mathcal{V}}\int_{0}^{\beta} d\tau\,e^{i\omega_n\tau} \braket{T_{\tau}\hat{\mathcal{J}}_{\alpha}(q, \tau)|\hat{\mathcal{J}}_{\beta}(-q, 0)}.\notag
\ee
where $\mathcal{V}$ is the volume of the system, and the current density operator $\hat{\mathcal{J}}$ is given by
\be
\hat{\mathcal{J}} = -\frac{\delta H}{\delta \vec{A}} = \pm e v_{F} \vec{\sigma}.\notag
\ee
Once we make the analytic continuation $i\omega_n\rightarrow \omega + i0^{+}$, the real frequency behavior is obtained easily. Thus, for each node we obtain
\small
\begin{align*}
\Pi_{\alpha\beta}(\omega) = \frac{e^2 v_F^2}{\mathcal{V}}\sum_{i, i', \vec{k}}\frac{f(\varepsilon_{i'}(\vec{k}))-f(\varepsilon_{i}(\vec{k}))}{\hbar\omega +\varepsilon_{i'}(\vec{k})-\varepsilon_{i}(\vec{k})+i0^{+}} \braket{\vec{k}i|\sigma_{\alpha}|\vec{k}i'}\braket{\vec{k}i'|\sigma_{\beta}|\vec{k}i}
\end{align*}
\normalsize
where $f(x)=1 \Big/ (1+e^{\beta x})$ is the Fermi function. The expression $H(\vec{k})\ket{\vec{k}i} = \varepsilon_i (\vec{k})\ket{\vec{k}i}$ with $i=1, 2$ gives the quasiparticle energies and eigenstates. We can evaluate the longitudinal and transverse polarizations $\Pi_{\alpha\beta}(\omega)$ when Fermi energy lies with nodes. Thus, in low frequency limit, longitudinal and transverse conductivities from both nodes are found to be expressions given in (\ref{conductivity1}) and (\ref{conductivity2}).

\subsection{D. Boundary Conditions }

Boundary conditions across the surface $\mathcal{S}$ between two regions of space are given by the statements: 1) Tangential component of electric field $\vec{E}$ is continuous across the interface, $\hat{n}\times (\vec{E}_1-\vec{E}_2) = 0$; 2) Normal component of magnetic field vector $\vec{B}$ is continuous, $\hat{n}\cdot (\vec{B}_1-\vec{B}_2) = 0$; 3) Normal component of electric flux density vector $\vec{D}$ is discontinuous by an amount equal to the surface current density, $\hat{n}\cdot (\vec{D}_1-\vec{D}_2) =\rho^s$; 4) Tangential component of the field $\vec{H}$ is discontinuous by an amount equal to the surface current density, $\hat{n}\times (\vec{H}_1-\vec{H}_2) = \vec{\cJ}^s$. Here $\hat{n}$ represents the unit normal vector to the surface $\mathcal{S}$ from region 2 to region 1. In our optical configuration, we do not have the third condition since there is no normal component of electric field. Therefore, we obtain the boundary conditions as in Table~\ref{table2},
\begin{table*}[ht]
\centering
{
    \begin{tabular}{@{\extracolsep{4pt}}ccl}
    \toprule
     \\
    $\mathbf{z}=0$  & {~~~~~} &
    $\begin{aligned}
    &\sum_{j=-}^{+}\,\left[A_1^{(j)}+C_1^{(j)}\right] = \sum_{j=-}^{+}\,\left[B_1^{(j)}+B_2^{(j)}\right],\\[3pt]
    &\sum_{j=-}^{+}\,j\left[A_1^{(j)}+C_1^{(j)}\right] = \sum_{j=-}^{+}\,j\left[B_1^{(j)}+B_2^{(j)}\right],\\[3pt]
    &\mu\sum_{j=-}^{+}\,j\left[\left(1 + 2\sigma_j\right)A_1^{(j)} - \left(1-2\sigma_j\right)C_1^{(j)}\right] = \sum_{j=-}^{+}\,j\tilde{\fn}_j\left[B_1^{(j)}-B_2^{(j)}\right],\\[3pt]
    &\mu\sum_{j=1}^{2}\,\left[A_1^{(j)}-C_1^{(j)}\right] = \sum_{j=-}^{+}\,\tilde{\fn}_j\left[B_1^{(j)}-B_2^{(j)}\right]. \\[3pt]
    \end{aligned}$\\
    \hline
    &\\[-10pt]
    $\mathbf{z}=1$  & {~~~~~} &$\begin{aligned}
    &\sum_{j=-}^{+}\,\left[A_2^{(j)}e^{i{\fK}}+C_2^{(j)}e^{-i{\fK}}\right] = \sum_{j=-}^{+}\,\left[B_1^{(j)}e^{i{\fK}_j}+B_2^{(j)}e^{-i{\fK}_j}\right],\\[3pt]
    &\sum_{j=-}^{+}\,j\left[A_2^{(j)}e^{i{\fK}}+C_2^{(j)}e^{-i{\fK}}\right] = \sum_{j=-}^{+}\,j\left[B_1^{(j)}e^{i{\fK}_j}+B_2^{(j)}e^{-i{\fK}_j}\right],\\[3pt]
    &\mu\sum_{j=-}^{+}\,j\left[\left(1+2\sigma_j\right)A_2^{(j)}e^{i{\fK}}-\left(1-2\sigma_j\right)C_2^{(j)}e^{-i{\fK}}\right] = \sum_{j=-}^{+}\,j\tilde{\fn}_j\left[B_1^{(j)}e^{i{\fK}_j}-B_2^{(j)}e^{-i{\fK}_j}\right],\\[3pt]
    &\mu\sum_{j=-}^{+}\,\left[A_2^{(j)}e^{i{\fK}}-C_2^{(j)}e^{-i{\fK}}\right] = \sum_{j=-}^{+}\,\tilde{\fn}_j\left[B_1^{(j)}e^{i{\fK}_j}-B_2^{(j)}e^{-i{\fK}_j}\right].
    \end{aligned}$\\[-8pt]
    &\\
    \hline
    \end{tabular}%
    }
    \caption{Boundary conditions for TE waves corresponding to TWS slab. Here effective indices $\tilde{\fn}_{\pm}$ are defined by (\ref{effectiveref}), and the quantity $\sigma_{j}$ is given in (\ref{ghfunc}).}
    \label{table2}
\end{table*}
In this table, we used ${\fK} := k_z L$ and ${\fK}_j := {\fK}\,\tilde{\fn}_j$. $\sigma_{j}$ is defined as follows for convenience
    \begin{align}
     \sigma_{j}:= \frac{\mu_0 \omega L}{2\fK} (\sigma_{yy} + ij \sigma_{yx}) = \frac{Z_0}{2\cos\theta} (\sigma_{yy} + ij \sigma_{yx}). \label{ghfunc}
    \end{align}
where $\sigma_{yy}$ and  $\sigma_{yx}$ are specified by Eqs.~\ref{conductivity1} and \ref{conductivity2}, and $j=+ $ or $-$ is implied.

\section*{Acknowledgement} We acknowledge the National Intelligence Academy of Türkiye for its valuable support during the execution of this
work and throughout the publication process. This work is funded by Scientific Research Projects Coordination Unit (BAP) of Istanbul University Project Number FBA-2020-35018.\\

\section*{Remarks}

\textbf{Contribution Statement:} All authors contributed equally to the whole work including calculations, analysis and manuscript typing.

\textbf{Data Availability Statement:}  All data generated or analysed during this study are included in this published article.

\textbf{Competing Interests:} The authors declare no competing interests.

\end{document}